\documentclass[longauth]{aa}
\usepackage{natbib}
\bibpunct{(}{)}{;}{a}{}{,} 
\usepackage{graphicx}
\usepackage{hyperref}
\usepackage{xcolor}
\usepackage{txfonts}

\usepackage{amsmath}	
\usepackage{orcidlink}
\usepackage{longtable}

\usepackage{soul}

\makeatletter
\renewcommand*\aa@pageof{, page \thepage{} of \pageref*{LastPage}}
\makeatother

\newcommand{\msolar}{M$_{\odot}$}
\newcommand\T{\rule{0pt}{2.6ex}}       
\newcommand\B{\rule[-1.2ex]{0pt}{0pt}} 

\begin{document} 

\title{JWST/MIRI detects the dusty SN1993J about 30 years after explosion}

\author{
\orcidlink{0000-0003-4610-1117}Tam\'as Szalai\inst{1,2} \and
\orcidlink{0000-0001-7473-4208}Szanna Zs{\'i}ros\inst{1} \and
\orcidlink{0000-0001-5754-4007}Jacob Jencson\inst{3} \and
\orcidlink{0000-0003-2238-1572}Ori D. Fox\inst{4} \and
\orcidlink{0000-0002-9301-5302}Melissa Shahbandeh\inst{3,4} \and
\orcidlink{0000-0002-9820-679X}Arkaprabha Sarangi\inst{5} \and
\orcidlink{0000-0001-7380-3144}Tea Temim\inst{6} \and
Ilse De Looze\inst{7} \and
\orcidlink{0000-0001-5510-2424}Nathan Smith\inst{8} \and 
\orcidlink{0000-0003-3460-0103}Alexei V. Filippenko\inst{9} \and
\orcidlink{0000-0001-9038-9950}Schuyler D.~Van Dyk\inst{10} \and
Jennifer Andrews\inst{11} \and
\orcidlink{0000-0002-5221-7557}Chris Ashall\inst{12} \and
\orcidlink{0000-0002-0141-7436}Geoffrey C. Clayton\inst{13} \and
Luc Dessart\inst{14} \and
Michael Dulude\inst{4} \and
Eli Dwek\inst{15,16} \and
\orcidlink{0000-0001-6395-6702}Sebastian Gomez\inst{4} \and
\orcidlink{0000-0001-5975-290X}Joel Johansson\inst{17} \and
\orcidlink{0000-0002-0763-3885}Dan Milisavljevic\inst{18,19} \and
Justin Pierel\inst{4} \and
\orcidlink{0000-0002-4410-5387}Armin Rest\inst{3,4} \and
\orcidlink{0000-0002-1481-4676}Samaporn Tinyanont\inst{20,21} \and
Thomas G. Brink\inst{9} \and
Kishalay De\inst{22} \and
Michael Engesser\inst{4} \and
Ryan J. Foley\inst{20} \and
Suvi Gezari\inst{4} \and
Mansi Kasliwal\inst{23} \and
Ryan Lau\inst{24} \and
Anthony Marston\inst{25} \and
Richard O'Steen\inst{4} \and
Matthew Siebert\inst{4} \and
Michael Skrutskie\inst{26} \and
Lou Strolger\inst{4} \and
Qinan Wang\inst{3} \and
Brian J. Williams\inst{27} \and
Robert Williams\inst{4} \and
Lin Xiao\inst{28,29} \and
WeiKang Zheng\inst{9}
}
     
\institute{
Department of Experimental Physics, Institute of Physics, University of Szeged, D{\'o}m t{\'e}r 9, 6720 Szeged, Hungary \\ 
\email{szaszi@titan.physx.u-szeged.hu}
\and
MTA-ELTE Lend\"ulet "Momentum" Milky Way Research Group, Szent Imre H. st. 112, 9700 Szombathely, Hungary 
\and
Department of Physics and Astronomy, Johns Hopkins University, Baltimore, MD 21218, USA 
\and
Space Telescope Science Institute, 3700 San Martin Drive, Baltimore, MD 21218, USA 
\and
DARK, Niels Bohr Institute, University of Copenhagen, Jagtvej 128, 2200 Copenhagen, Denmark 
\and
Department of Astrophysical Sciences, Princeton University, Princeton, NJ 08544, USA 
\and
Sterrenkundig Observatorium, Ghent University, Krijgslaan 281 -- S9, 9000 Gent, Belgium 
\and
Steward Observatory, University of Arizona, 933 N. Cherry St, Tucson, AZ 85721, USA 
\and
Department of Astronomy, University of California, Berkeley, CA 94720-3411, USA 
\and
Caltech/IPAC, Mailcode 100-22, Pasadena, CA 91125, USA 
\and
Gemini Observatory, 670 N. Aohoku Place, Hilo, Hawaii, 96720, USA 
\and
Department of Physics, Virginia Tech, Blacksburg, VA 24061, USA 
\and
Department of Physics \& Astronomy, Louisiana State University, Baton Rouge, LA, 70803 USA  
\and
Institut d’Astrophysique de Paris, CNRS–Sorbonne Université, 98 bis boulevard Arago, F-75014 Paris, France 
\and
NASA at Goddard Space Flight Center, Code 665, Greenbelt, MD 20771, USA 
\and
Center for Astrophysics | Harvard \& Smithsonian, 60 Garden Street, Cambridge, MA 02138-1516, USA 
\and
Oskar Klein Centre, Department of Physics, Stockholm University, AlbaNova, SE-10691 Stockholm, Sweden 
\and
Purdue University, Department of Physics and Astronomy, 525 Northwestern Ave, West Lafayette, IN 47907, USA 
\and
Integrative Data Science Initiative, Purdue University, West Lafayette, IN 47907, USA 
\and
Department of Astronomy and Astrophysics, University of California, Santa Cruz, CA 95064, USA 
\and
National Astronomical Research Institute of Thailand, 260 Moo 4, Donkaew, Maerim, Chiang Mai, 50180, Thailand 
\and
MIT-Kavli Institute for Astrophysics and Space Research, 77 Massachusetts Ave., Cambridge, MA 02139, USA 
\and
Cahill Center for Astrophysics, California Institute of Technology, 1200 E. California Blvd. Pasadena, CA 91125, USA 
\and
NSF’s NOIRLab, 950 N. Cherry Avenue, Tucson, 85719, AZ, USA 
\and
European Space Agency (ESA), ESAC, 28692 Villanueva de la Canada, Madrid, Spain 
\and
Department of Astronomy, University of Virginia, Charlottesville, VA 22904-4325, USA 
\and
X-ray Astrophysics Laboratory, NASA / Goddard Space Flight Center (GSFC), Greenbelt, MD 20771, USA 
\and
Department of Physics, College of Physical Sciences and Technology, Hebei University, Wusidong Road 180, Baoding 071002, China 
\and
Department of Physics, College of Physical Sciences and Technology, Hebei University, Wusidong Road 180, Baoding 071002, China 
}

\date{Accepted XXX. Received YYY; in original form ZZZ}

  \abstract
   {Core-collapse supernovae (CCSNe) have long been considered to contribute significantly to the cosmic dust budget. New dust cools quickly and is therefore detectable at mid-infrared (mid-IR) wavelengths. However, before the era of the {\it James Webb Space Telescope} ({\it JWST}), direct observational evidence for dust condensation was found in only a handful of nearby CCSNe, and dust masses ($\sim 10^{-2}-10^{-3}$ \msolar, generally limited to $< 5$~ yr and to $> 500$~K temperatures) have been 2--3 orders of magnitude smaller than either theoretical predictions or dust amounts found by far-IR/submm observations of Galactic SN remnants and in the very nearby SN~1987A.}
   {As recently demonstrated, the combined angular resolution and mid-IR sensitivity of {\it JWST} finally allow us to reveal hidden cool ($\sim 100$--200~K) dust reservoirs in extragalactic SNe beyond SN 1987A. Our team received {\it JWST}/MIRI time 
   for studying a larger sample of CCSNe to fill the currently existing gap in their dust formation histories. The first observed target of this program is the well-known Type IIb SN~1993J appeared in M81.}
   {We generated its spectral energy distribution (SED) from the current {\it JWST}/MIRI F770W, F1000W, F1500W, and F2100W fluxes. We fit single- and two-component silicate and carbonaceous dust models to the SED in order to determine the dust parameters.}
   {We found that SN~1993J still contains a significant amount ($\sim 0.01$~\msolar) of dust $\sim 30$~yr after explosion. Comparing these results to those of the analysis of earlier {\it Spitzer Space Telescope} data, we see a similar amount of dust now that was detected $\sim 15$--20~yr ago, but at a lower temperature (noting that the modeling results of the earlier {\it Spitzer} SEDs have strong limitations). We also find residual background emission near the SN site (after point-spread-function  subtraction on the {\it JWST}/MIRI images) that may plausibly be attributed to an IR echo from more distant interstellar dust grains heated by the SN shock-breakout luminosity or ongoing star formation in the local environment.
   }
 {}
   \maketitle
\section{Introduction}

Core-collapse supernovae (CCSNe), the energetic final explosions of evolved massive ($\gtrsim 8$~\msolar) stars, offer unique possibilities to (i) study extreme physical processes, (ii) uncover details about pre-explosion stellar evolution, and (iii) measure cosmic distances.
%
While most CCSNe fade over the course of several months to years, the {\it Spitzer Space Telescope} ({\it Spitzer}) ``Warm'' (post-cryogenic) mission (i.e., 3.6 and 4.5~$\mu$m) highlighted a subset of dusty SNe that can remain bright for many years, even decades, post-explosion \citep[e.g.,][]{fox10,fox11,fox13,tinyanont16,szalai19,szalai21}. The increased sensitivity of the {\it James Webb Space Telescope} ({\it JWST}) and its access to even longer wavelengths (e.g., 25~$\mu$m) has resurrected the transient community's interest in such dusty SNe. The origin and heating mechanism of the dust can have several important implications. If the dust is newly formed in the ejecta, the inferred dust masses could provide the much sought-after evidence supporting SNe as significant sources of dust in the Universe \citep[e.g.,][]{dwek07}. If there is also or only pre-existing dust in the circumstellar medium (CSM) at the time of explosion, it can be used as a proxy to trace the pre-SN mass-loss history and constrain the progenitor system \citep[e.g.,][]{fox11}.

Up to the beginning of the {\it JWST} era, there have been only a few objects also observed at longer mid-IR wavelengths: several Type II-P SNe up to a few years old \citep[e.g.,][]{kotak09,fabbri11,meikle11,szalai11,szalai13}, as well as older objects like SNe~1978K \citep{tanaka12}, 1980K \citep{sugerman12}, or 1995N \cite{vandyk13}. However, only
the famous, nearby SN~1987A in the Large Magellanic Cloud was possible to study in detail not just by {\it Spitzer}  \citep{bouchet06,dwek10,arendt20}, but also in the far-IR/submm regime \citep[via {\it Herschel} and ALMA, see][]{matsuura11,matsuura19,indebetouw14}. 
Thus, for the most dusty extragalactic SNe, we have no information on temperatures $\lesssim 150$~K, where the bulk of the dust is thought to reside --- as was also shown by recent far-IR and submm observations of old Galactic SN remnants (SNRs) like Cassiopeia~A \citep{barlow10,sibthorpe10,arendt14} and the Crab \citep{gomez12,temim13,delooze19}. A number of questions remain and the phase space of such observations (in terms of mass/temperature of dust vs. SN age) remains relatively unpopulated.
Moreover, it should be noted that while {\it Spitzer} had  good mid-IR sensitivity, its angular resolution was too poor to separate most extragalactic SNe from their nearby host-galaxy emission from star-forming regions, \ion{H}{ii} regions, etc. (especially at $\lambda \gtrsim$ 20~$\mu$m).

%

%
{\it JWST} offers a new opportunity to detect the late phases of cool ($\sim 100$--200~K) dust in extragalactic SNe beyond SN~1987A. 
{\it JWST} has the potential to detect (i) cooler dust grains at wavelengths $>4.5$~$\mu$m, (ii) the 10~$\mu$m silicate feature that can distinguish grain compositions, and (iii) faint emission from the SN at very late epochs that would have gone undetected by {\it Spitzer} and any other mid-IR spacecraft.
Only a few months after the start of its scientific mission, {\it JWST} has achieved important results in this field. Our team has already detected a significant amount of cool dust in SNe~IIP 2004et and 2017eaw \citep{Shahbandeh2023}, as well as in SN~IIL 1980K \citep{Zsiros_2024}. In SN~2004et, the observations have uncovered the largest newly formed ejecta dust masses in an extragalactic SN other than SN~1987A, with $\gtrsim 10^{-2}$~\msolar\ of dust at a temperature of $\sim 140$~K. 


In this paper, we present {\it JWST} observations of another nearby, famous event, SN~1993J. This object is the prototype of Type IIb explosions, which form a transitional group between H-rich Type II and H-free Type Ib/c CCSNe \citep[see, e.g.,][]{filippenko88,Filippenko_97,filippenko93,nomoto93} and constitute about 10\% of core-collapse SNe \citep{Smith_2011}.
Owing to its proximity \citep[$3.63 \pm 0.31$~Mpc;][]{freedman01} and its fortunate location in the outskirts of the host galaxy, M81, SN~1993J has become one of the best-observed SNe, possessing various long-term multiwavelength datasets and detailed analyses published in the literature. 
We also know much about its progenitor system. The exploding star, likely a K0-type red supergiant (RSG), was directly identified in pre-explosion images \citep{aldering94,cohen95}, and had disappeared years after explosion \citep{maund09}. The progenitor may have been a member of a massive binary system \citep{maund04} with a hotter (B2 type) companion that likely was directly detected \citep{fox14}. 

SN~1993J also showed various signs of ongoing interaction between the shock and the circumstellar material (CSM) from early to late phases in the optical, radio, and X-ray bands  \citep[e.g.,][]{matheson00,weiler07,chandra09,smith17}; see \citet[][hereafter Z22]{zsiros22} for a recent review.
Note that SN~1993J is one of the very few extragalactic SNe besides SN 1987A where, via very-long-baseline radio interferometry, it was possible to resolve how the expanding SN shock is running into the CSM \citep{bietenholz01,bietenholz03,bartel02}.
Signs of dust formation have been  published based on early-time near-IR observations \citep{matthews02} and late-time optical spectral analysis \citep{smith17,bevan17}. 
In Z22, some of us presented a detailed analysis of the complete {\it Spitzer} mid-IR light curve (LC) and spectral energy distribution (SED), and  modeling of the object \citep[part of the {\it Spitzer} LC was also published by][]{tinyanont16}. 

With the recent {\it JWST} data, SN~1993J becomes one of the few SNe for which a dust-formation history can be followed during the first decades after explosion.
In Section \ref{sec:obs}, we describe the steps of data reduction, while  Section \ref{sec:anal} presents the results of our analysis based on modeling the recent mid-IR SED. We discuss our findings in Section \ref{sec:disc}
and summarize our conclusions in Section \ref{sec:conc}.

\section{Observations and data reduction}\label{sec:obs}


\begin{figure}
        \includegraphics[width=\columnwidth]{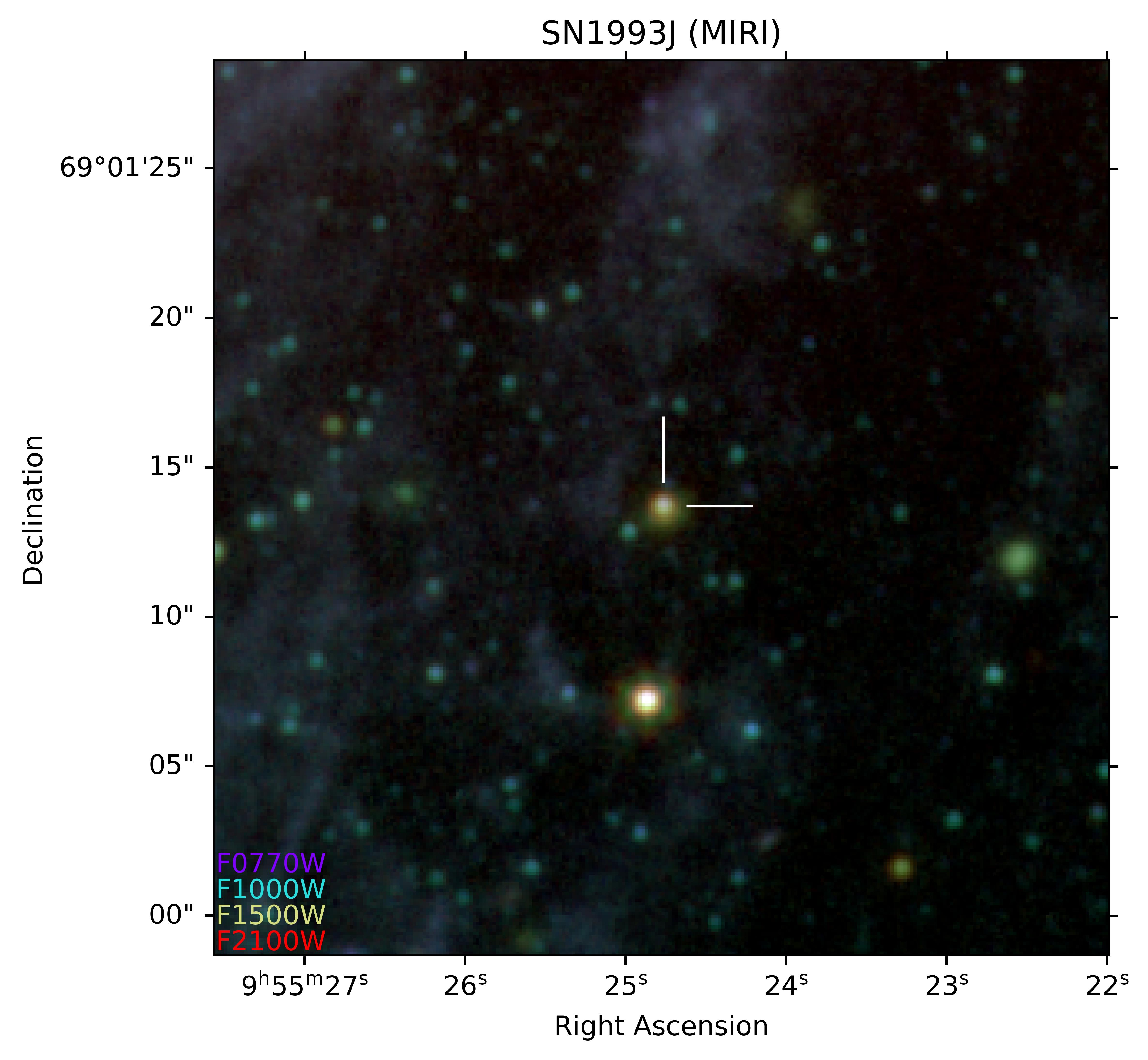}
    \caption{{\it JWST}/MIRI composite image of SN~1993J [$\alpha(2000) = 09^{\rm hr}55^{\rm m}24.778^{\rm s}$, $\delta(2000) = +69^{\circ} 01\arcmin 13\farcs 70$] obtained on 2024 Feb. 13.6 UTC (11,280 days post-explosion); white tick marks show the position of the target.} 
    \label{fig:93J_color}
\end{figure}

{\it JWST} observing program \#3921 (PI O.~D. Fox) \footnote{\href{https://www.stsci.edu/jwst/science-execution/program-information?id=3921}{https://www.stsci.edu/jwst/science-execution/program-information?id=3921}} is designed as a SURVEY to image a large, diverse sample of SNe with the Mid-Infrared Instrument (MIRI; \citealt{Bouchet_2015, Ressler_2015, Rieke_2015, Rieke_2022}).
The observations are being acquired in the F770W, F1000W, F1500W, and F2100W filter bands, using the FASTR1 readout pattern in the FULL array mode and a 4-point extended source dither pattern.
A description of our detailed calibration process of the {\it JWST}/MIRI images was recently published by \citet{Shahbandeh2023}. We use the JWST HST Alignment Tool \citep[JHAT;][]{Rest_2023} to align {\it JWST} and {\it HST} images of the fields (when available) with each other.

The first observed target of the SURVEY 3921 program was SN~1993J; MIRI images were obtained on 2024 Feb. 13.6 UTC (11,280 days after explosion, $t_0$ = 49074.0 MJD, \citealt{lewis94}).
The SN can be identified as a clear and bright point source at all wavelengths from 7.7 to 21.0~$\mu$m; see a composite image of the field in Fig. \ref{fig:93J_color}.

To measure the fluxes of SN~1993J in {\it JWST}/MIRI images, we followed the method described in detail by \citet{Shahbandeh2023}. 
We performed point-spread-function (PSF) photometry on background-subtracted level-two data products using \texttt{WebbPSF} \citep{Perrin_2014} implemented in the \texttt{space-phot} package\footnote{https://zenodo.org/records/12100100} \citep{pierel_2024_12100100}. 
We experimented with multiple PSF sizes
in the fitting that variably resulted in underestimation or overestimation of the local background emission. The resulting fluxes of all four dithers of each filter were then averaged, and we incorporate the variations seen from using multiple PSF sizes into our estimates of the measurement uncertainties. We examine the local background emission from the PSF-fitting residuals in more detail in Section~\ref{sec:anal_echo}. The total (Galactic + host) reddening value of $E(B-V)$ = 0.19$\pm$0.09 mag \citep{richardson06} implies that the extinction is negligible in the mid-IR range. The final results of our {\it JWST}/MIRI photometry of SN~1993J are presented in Table~\ref{tab:JWST_phot}.

\begin{table}
	\centering
	\caption{JWST/MIRI AB magnitudes and fluxes of SN~1993J 
 on 2024~Feb.~13.6 UTC (day 11,280 post-explosion).}
	\label{tab:JWST_phot}
    \renewcommand{\arraystretch}{1.5}
	\begin{tabular}{llll} 
		\hline
		Filter & AB mag & $F_{\nu}$ & $F_{\lambda}$ \\
        & & ($\mu$Jy) & erg s$^{-1}$ cm$^{-2}$ \AA$^{-1}$ \\
        \hline
		F770W & 20.52$\pm$0.04 & 22.5$\pm$1.0 & (1.14$\pm$0.04)$\times 10^{-19}$\\
        F1000W & 19.88$\pm$0.05 & 40.6$\pm$1.9 & (1.22$\pm$0.06)$\times 10^{-19}$\\
        F1500W & 18.39$\pm$0.18 & 160.0$\pm$26.5 & (2.13$\pm$0.35)$\times 10^{-19}$\\
        F2100W & 17.43$\pm$0.06 & 387.3$\pm$21.4 & (2.63$\pm$0.15)$\times 10^{-19}$\\
		\hline
	\end{tabular}
    \end{table}

\begin{figure}
    \includegraphics[width=\columnwidth]{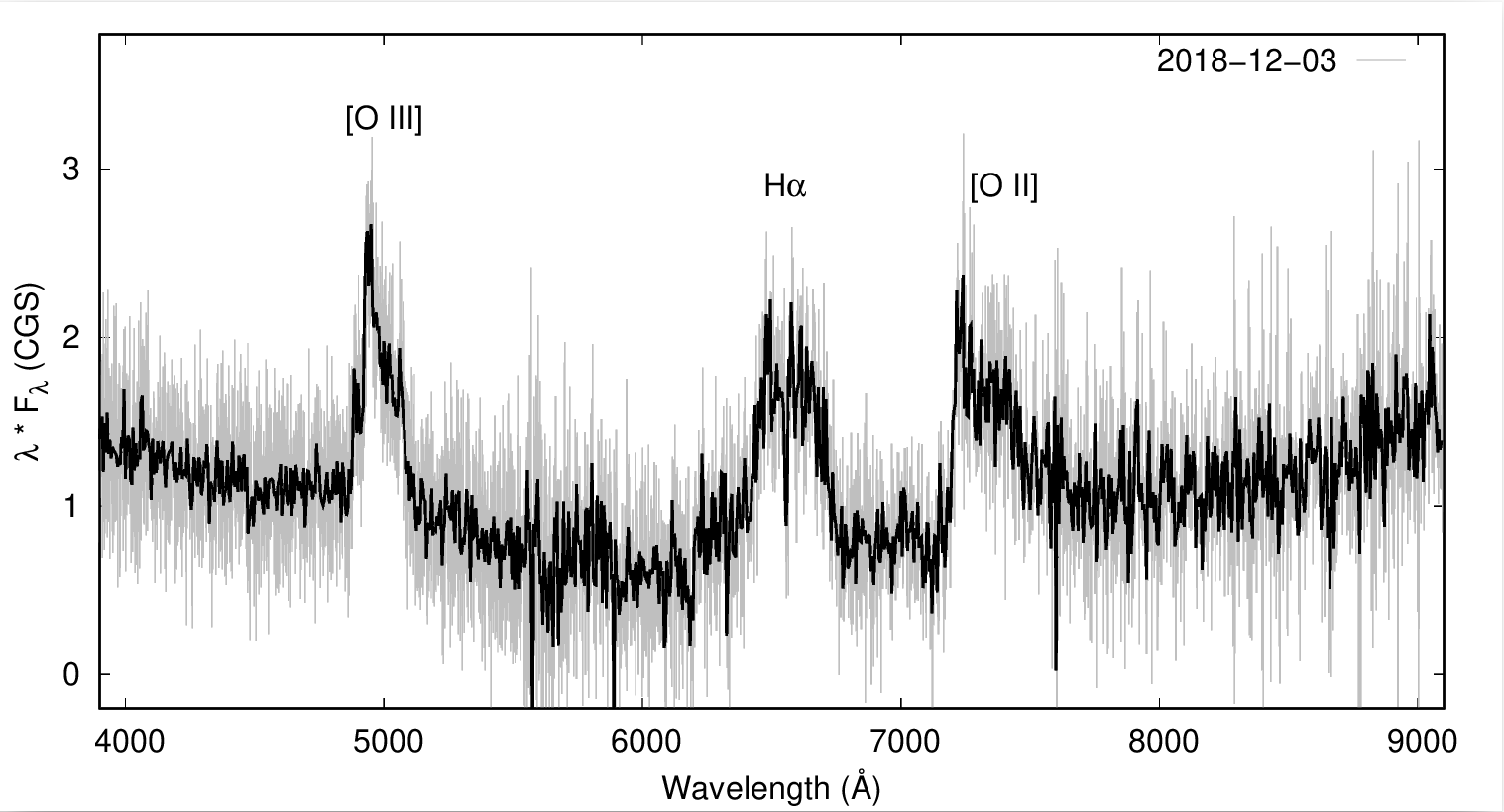}
    \caption{High-resolution optical spectrum of SN~1993J obtained with Keck/LRIS on 2018-12-03, 9381 days after explosion; dark solid line marks the 5 \AA-binned data.}
    \label{fig:93J_spec}
\end{figure}

During the analysis described in Sect. \ref{sec:anal_multi}, we also used a single unpublished late-time optical spectrum of SN~1993J obtained with the Keck Low Resolution Imaging Spectrometer (LRIS; \citealp{Oke_1995}) on 2018 Dec. 03 (at epoch 9381 days), see Fig. \ref{fig:93J_spec}.
The spectrum was acquired with the slit oriented at or near the parallactic angle to minimise slit losses caused by atmospheric dispersion \citep{Filippenko_1982}. 
The LRIS observation utilised the $1\arcsec$-wide slit, 600/4000 grism, and 400/8500 grating to produce a similar spectral resolving power ($R \approx 700$--1200) in the red and blue channels. 
Data reduction followed standard techniques for CCD processing and spectrum extraction using the LPipe data-reduction pipeline \citep{Perley_2019}. 
Low-order polynomial fits to comparison-lamp spectra were used to calibrate the wavelength scale, and small adjustments derived from night-sky lines in the target frames were applied. 
The spectrum was flux calibrated using observations of appropriate spectrophotometric standard stars observed on the same night, at similar airmasses, and with an identical instrument configuration.

\section{Analysis}\label{sec:anal}

\subsection{Modeling of the mid-IR SEDs of SN~1993J}\label{sec:anal_sed}

Mid-IR (continuum) excess relative to the Rayleigh-Jeans tail in the late-time SED of an SN typically indicates the
presence of dust. As described above, this dust may be (i) newly formed, in either the inner unshocked ejecta or in the cool dense shell (CDS) of post-shocked gas lying between the forward and reverse shocks, and/or (ii) pre-existing, formed in a steady wind from the progenitor or during a short-duration pre-SN outburst. In either case, different heating mechanisms, geometries, grain composition and size distribution, and dust clumpiness effects should be taken into account to find a proper description of the physical background of the observed mid-IR radiation; see  more details provided by for example, \citet[][and references therein]{Shahbandeh2023}.

A detailed analysis of the properties of the dust content and its possible origin and heating effects is described by Z22, based on the full {\it Spitzer} dataset of the event obtained between 2003 and 2019 (10--26~yr after explosion). 
The conclusion of this study was that all the dust suggested by {\it Spitzer} data can be newly formed and is located in the inner ejecta and/or in the CDS; this picture is also strengthened by the modeling of red-blue line-profile asymmetries found in late-time optical spectra of the object \citep[][]{bevan17,smith17}.
At the same time,  the results of Z22 also allow for the presence of pre-existing dust, heated collisionally by hot gas in the reverse shock (assuming a possible range of 5000--15,000 km s$^{-1}$ for the shock velocity), or (more likely) radiatively by energetic photons from shock-CSM interaction.

\begin{figure}
    \includegraphics[width=\columnwidth]{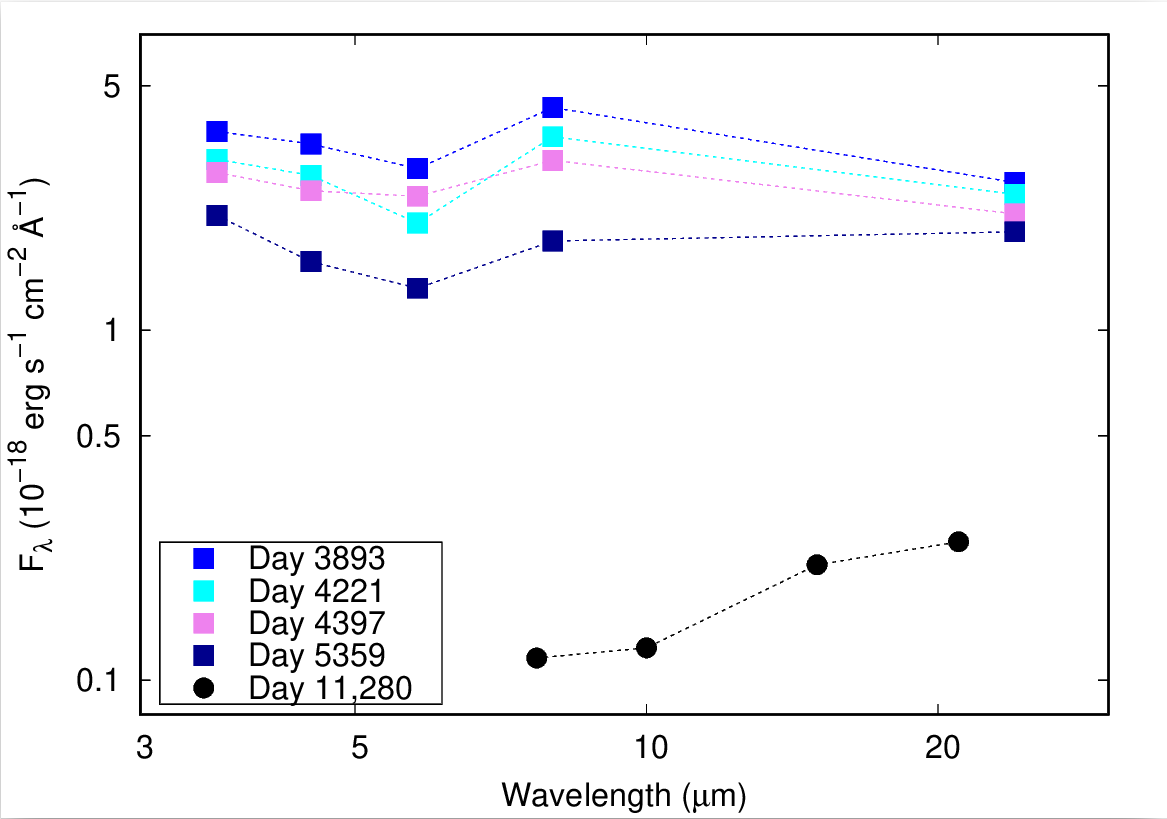}
    \caption{Evolution of observed mid-IR SEDs of SN~1993J between 2003 and 2024. Color squares and black circles denote {\it Spitzer}/IRAC+MIPS \citep{zsiros22} and {\it JWST}/MIRI data points, respectively.}
    \label{fig:93J_SEDs}
\end{figure}

\begin{figure}
    \includegraphics[width=\columnwidth]{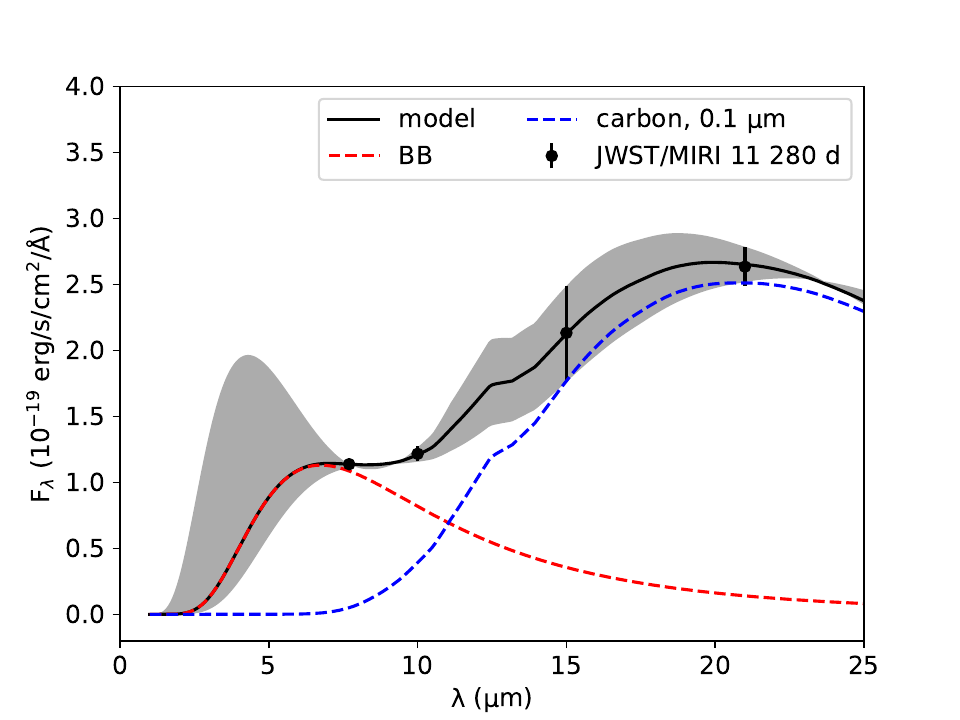}
    \caption{
    Best-fit two-component --- cold ($T_\textrm{cold} \approx 120$~K) amorphous carbon dust + hot ($T_\textrm{hot} \approx 430$~K) BB --- model fitted to the four-point {\it JWST}/MIRI SED of SN~1993J (day 11,280). Gray regions pertain to models fitted to the upper and lower constraints on the fluxes based on the photometric uncertainties, respectively. Model parameters are shown in Table \ref{tab:dustpar}. Filled circles denote measured fluxes. The hot component was replaced also by a fixed $T = 1000$~K BB and a power-law function, causing no significant changes in the parameters of the cold dust component (see details in the text).}
    \label{fig:93J_SED_models}
\end{figure}

Here we do not repeat all the steps presented by Z22 but examine (i) the validity of the conclusions of the previous work, and (ii) the points where new {\it JWST} data allow us to carry out an even more detailed analysis.

First, we compare the fluxes of SN~1993J measured by {\it JWST}/MIRI (day 11,280 post-explosion) to those measured by {\it Spitzer} (between days 3893 and 5359) and published by Z22. The evolution of the mid-IR SEDs of the object between days 3893 and 11,280 is shown in Fig. \ref{fig:93J_SEDs}. {\it JWST} fluxes are an order of magnitude lower than the earlier ones and the shape of the SED seems to shift toward lower temperatures. Note that there are further {\it Spitzer}/IRAC 3.6 and 4.5~$\mu$m data up to $\sim 9650$ days; the SN shows continuously decreasing fluxes at both wavelengths \citep[see][and Z22]{tinyanont16}.
We also note that recent {\it JWST} imaging strengthens our previous assumption that the environment of SN~1993J is relatively smooth with a low local IR background (see Fig. \ref{fig:93J_color}). It gives further justification for the reliability of our previously published {\it Spitzer} photometry (there was no possibility for subtracting pre-explosion images; see Z22).

Next, we carry out a simple analytic modeling of the {\it JWST}/MIRI SED of SN~1993J described by 
\cite{hildebrand83} (similarly to the work of Z22 on {\it Spitzer} SEDs).
This method assumes only the thermal emission of an optically thin dusty shell at a single equilibrium temperature $T_d$, with a dust mass of $M_d$ and particle radius of $a$. 
The observed flux of the dust can then be written as

\begin{equation}
    F_{\lambda} = M_d B_{\lambda}(T_d) \kappa(a)/D^2\, ,
    \label{eq:flux}
\end{equation}

\noindent 
where $B_{\lambda}(T_d)$ is the Planck function, $\kappa(a)$ is the dust mass absorption coefficient, and $D$ is the distance to the source. 

For building the model SEDs, we calculated filter-integrated fluxes using Eq.~\ref{eq:flux} convolved with the {\it JWST}/MIRI filter transmission profiles and fit them to the {\it JWST} data. 
Note that, while our group \citep{Shahbandeh2023} adopted a more general formalism from \citet{Dwek_2019} allowing for the presence of optically thick dust, we focus here only on the optically thin case because of the decades-long age of the SNR \citep[just as we did in the case of SN~1980K;][]{Zsiros_2024}.
We also note that using only Eq. \ref{eq:flux} does not allow us to take into account various geometries or clumpiness factors; this avoids the overinterpretation of a four-point SED and leads us rather to draw conclusions based on the parameters determined from the simplest assumptions. Nevertheless, we must handle all the constraints of this model carefully.
%
%
%

From the previously published {\it Spitzer} dataset, it was difficult to infer the dust composition in SN~1993J. 
These data do not cover the 8--24~$\mu$m range (except for a single, noisy {\it Spitzer}/IRS spectrum obtained in 2008), which could have been critical for disentangling the spectral features of Si-rich and carbonaceous dust. Thus, in Z22, both amorphous carbon and silicate dust models were fitted to the {\it Spitzer} SEDs.
In the case of the current {\it JWST} dataset, fluxes measured with the F1000W, F1500W, and F2100W filters could be more helpful in revealing the true dust composition. 
%
%

Following \citet{Shahbandeh2023}, the absorption and emission properties for amorphous carbon (amC) and silicate grains are obtained from \cite{Zubko_2004} and \cite{Draine_2007}, respectively (see \citealp{sarangi22} for the values of absorption coefficients $\kappa_{\lambda}$).
As a first step,
we applied a 0.1~$\mu$m grain size, just as we did before in our previous {\it JWST/MIRI} data papers \citep{Shahbandeh2023,Zsiros_2024} (and as also did by Z22 for the {\it Spitzer} data analysis of SN~1993J). 
We found that the four-point day-11,280 {\it JWST}/MIRI SED of SN~1993J cannot be properly fitted with any single-component amC or silicate dust models.
While the measured 7.7 $\mu$m flux appears to show an excess with respect to each single-component model, we added a hot blackbody (BB) component and obtained an adequate fit assuming a cold amC dust component associated with a hot BB (see Fig. \ref{fig:93J_SED_models}). $a=0.1 \mu$m silicate models do not result in reasonable fits even in the two-component cases.

Following our referee's advice, we also performed the SED fitting assuming larger ($a$=1.0 and 5.0 $\mu$m) grain radii. We found that even this choice does not allow us to find good single-component solutions; nevertheless, two-component models work if we use $a$=1.0/5.0 $\mu$m amC or $a$=5.0 $\mu$m silicate dust associated with a hot BB (see Fig. \ref{fig:SED_large_grains} in Appendix \ref{sec:appendixA}).
Note, however, that such large grains are not expected in SNe at this phase (especially in SNe IIb, see in Sect. \ref{sec:disc}); furthermore, this dataset does not allow us to truly disentangle models with different grain-size distributions. Thus, from now on, we refer to the results of the original fit assuming a cold $a$=0.1 $\mu$m amC dust + a hot BB. 


Since fitting a two-component model to only four SED points 
leads to unphysically low parameter uncertainties, we repeated the fit based on the upper and lower constraints placed on the {\it JWST}/MIRI fluxes by the photometric uncertainties in Table \ref{tab:JWST_phot}.
Thus, the cold dust component has an inferred mass of $M_\textrm{cold}$ = (8.9$^{+4.5}_{-2.3}$) $\times 10^{-3}$~M$_{\odot}$ and a temperature of $T_\textrm{cold} = 116\pm10$~K. 
For the average flux values, we also calculated the luminosity and the BB radius of the cold components (i.e., the minimum size of an optically thin, spherical dust-forming region), finding $L_\textrm{cold} \approx1.0 \times 10^{38}$~erg~s$^{-1}$ and $R_\textrm{BB,cold} \approx 1.3 \times 10^{16}$~cm. As Fig. \ref{fig:93J_SEDs}  already suggested, the current value of $L_\textrm{cold}$ is well below those measured from {\it Spitzer} SEDs from $\gtrsim 5$ yr prior [(10.0--15.6) $\times 10^{38}$~erg~s$^{-1}$; see Z22].

As can be recognized from Fig. \ref{fig:93J_SED_models}, how the assumed hot component affects the parameters of the cold dust component can be an important question here.
A similar hot gas/dust component has been previously detected in other decades-old SNe such as SN~1980K \citep{Zsiros_2024}, SN~1987A \citep{dwek10,arendt16,arendt20}, and SN~2004et \citep{Shahbandeh2023}. The general assumption is ongoing circumstellar interaction in the close environment of these SNe, and this is the case for SN~1993J as well: 
as mentioned above, SN~1993J has shown various signs of CSM interaction 
and, based on long-term monitoring of its optical spectra, this level remains strong even $\sim 20$~yr after explosion \citep[see][and Fig. \ref{fig:93J_lum} of this work]{milisavljevic12,smith17}.

Nevertheless, based on {\it JWST} data alone, we cannot conclusively determine the current temperature of the hot component. Unfortunately, we are not aware of any near-IR data from the past few years (the last {\it Spitzer} data at 3.6 and 4.5~$\mu$m were obtained in 2019; note, however, that the SN was then very faint at these wavelengths, so we cannot follow the real steepness of its decline after that).
If we fit our two-component model using free parameters, we get $T_\textrm{hot} = 430^{+150}_{-50}$K for the temperature of the hot component. Between days 3893 and 5359, we do not see a clear evolution of $T_\textrm{hot}$ (its value varies between 640 and 780~K; see Z22); note, however, that these SEDs are also unconstrained below 3.6~$\mu$m. Thus, following the method described by \citet{Zsiros_2024}, we repeated the fitting procedure, fixing $T_\textrm{hot} = 1000$~K to see how it affects the parameters of the cold dust component. As shown in Table \ref{tab:dustpar}, these differences are essentially small. Since the nature of the hot component is not necessarily thermal, we also run a test assuming that the hot component has a power-law nature ($ F_{\lambda} \propto \lambda^{0.3}$, fitted to the day 5359 3.6--5.8~$\mu$m {\it Spitzer} fluxes). However, in this case, we see that this function declines quickly below $\sim 10^{20}$~erg~s$^{-1}$cm~$^{-2}~\AA^{-1}$ beyond 10~$\mu$m; thus, its influence on the cold dust component is even smaller than that of the tested hot dust/BB components.

We note that Z22 used older absorption coefficients during the modeling of {\it Spitzer} SEDs. Hence, for an improved comparison, we also repeated the fitting of two-component amC dust models to the last (day 5359) {\it Spitzer} SED using the $\kappa$ values adopted from \cite{sarangi22}.
Results of this repeated fitting are  shown in Table \ref{tab:dustpar}. We discuss in detail the evolution of dust parameters in Section \ref{sec:disc}.

Note, furthermore, that we examined the potential contribution of the long-wavelength synchrotron radiation to the mid-IR SED (emerging from shock-CSM interaction), just as done recently by \cite{Larsson_2023} and \cite{Jones_2023} in the case of SN~1987A. Since SN~1993J has also been a target of long-term radio observations, we followed their method and extrapolated the radio LC model parameters of SN~1993J from \cite{weiler07} (see their Eq. (1), including both nonthermal synchrotron self-absorption and thermal free-free absorbing components) to the mid-IR range at the epoch of 11,280 days. We found that the original model values may give a $\lesssim$10 $\mu$Jy contribution at 25.5 $\mu$m and even lower amounts at shorter wavelengths. Also note that the real contribution may be even  smaller, since the general model of \cite{weiler07}, as the authors discuss, gives an overestimation of the measured radio fluxes after $\sim 3000$ days \citep[the achromatic break in the radio LCs of SN~1993J was also described by, e.g.,][]{martividal11,kundu19}. So, unlike in the case of SN~1987A, the contribution of the synchrotron emission to the very-late-time mid-IR SED of SN~1993J seems to be negligible.


\begin{table*}
	\centering
	\caption{Parameters of the best-fit two-component (cold amorphous carbon dust + hot blackbody) models for the day 5359 {\it Spitzer} and the day 11,280 {\it JWST}/MIRI SED of SN~1993J, both calculated using $\kappa$ files from \citet{sarangi22}.} 
	\label{tab:dustpar}
	\begin{tabular}{ll|cc|cc|c} 
		\hline
        \hline
		Data & Epoch & $T_\textrm{cold}$ & $M_\textrm{cold}$ & $L_\textrm{cold}$ & $R_\textrm{BB,cold}$ & $T_\textrm{BB,warm}$\T \\
       & (days) & (K) & ($10^{-3}~{\rm M}_{\odot}$) & (10$^{38}$ erg s$^{-1}$) & (10$^{16}$ cm) & (K) \B\\
        \hline
        {\it Spitzer} (3.6--8.0 + 24 $\mu$m) & 5359 & 167$\pm$4 & 14.3$\pm$2.9 & 10.0 & 2.9 & 901$\pm$52 \T\B\\
        \hline
        {\it JWST} (7.7, 10.0, 15.0, 21.0 $\mu$m) & 11,280 & 116$\pm$10 & 8.9$^{+4.5}_{-2.3}$ & 1.0 & 1.3 & 430$^{+150}_{-50}$ \T \B\\
        ~ & ~ & 128$\pm$2 & 5.2$\pm$0.5 & -- & -- & 1000 (fixed) \B\\
		\hline
	\end{tabular}
\end{table*}

\subsection{Comparison of late-time multiwavelength data}\label{sec:anal_multi}
It is also worth comparing the mid-IR evolution of SN~1993J to published late-time UV/optical flux changes \citep[see e.g.][]{fransson02,Dessart_2022,Dessart_2023a}. This step could help reveal the role of CSM interaction in the potential heating of the dust content of the SN shown by recent {\it JWST} measurements.

First, we examine the late-time {\it HST} photometry of the object. Based on the results presented by \cite{Baer-Way_2024} --- which also contain  previously published data from \cite{vandyk02} and \cite{fox14} --- the F336W fluxes show a moderate decline by 0.026 and 0.036~mag~(100~d)$^{-1}$ in the ranges 6903--8022~d and 8022--10,123~d, respectively. In the F814W filter, there is a steeper decline of 0.073~mag~(100~d)$^{-1}$ during days 6903--10,123, preceded by a slower decrease of 0.029~mag~(100~d)$^{-1}$ during days 2990--6903 (note, however, that these decline rates are determined from single pairs of data points).

We can also do a similar analysis of previously published, well-sampled {\it Spitzer}/IRAC LCs (\citealt{tinyanont16}; Z22). While there are 3.6 and 4.5~$\mu$m data up to $\sim 9600$ days,
the object seems to fade into the background at these wavelengths after $\sim 5500$ days. Thus, we only use days 3875--5345 data here, and determine a decline of 0.039, 0.057, 0.060, and 0.071~mag~ (100~d)$^{-1}$ for 3.6, 4.5, 5.8, and 8.0~$\mu$m {\it Spitzer} LCs, respectively. Thus, while {\it Spitzer} data have quite large uncertainties (0.2--0.3 mag), it seems that fading is slower at shorter wavelengths (and the decline rate of {\it HST} F814W data during days 2990--6903 also fits  this trend; see \citealt{Baer-Way_2024}). This seems to further strengthen the importance of the ongoing CSM interaction in the environment of SN~1993J. 

Using the data published by Z22 and the decline rates reported above, we also did an extrapolation of the 8.0~$\mu$m {\it Spitzer}/IRAC LC to day 11,280 (17.25 mag) and compared to the measured {\it JWST}/MIRI 7.7~$\mu$m brightness (16.19 mag, both values in Vega magnitudes). While the difference between the spatial resolution of the two telescopes and the (slightly) different transmission curves of the two filters should make us careful, we suggest that the $\sim 1$ mag difference really implies a slowing decline rate after day 5345. 

\begin{figure*}
\centering
    \includegraphics[width=0.4\textwidth]{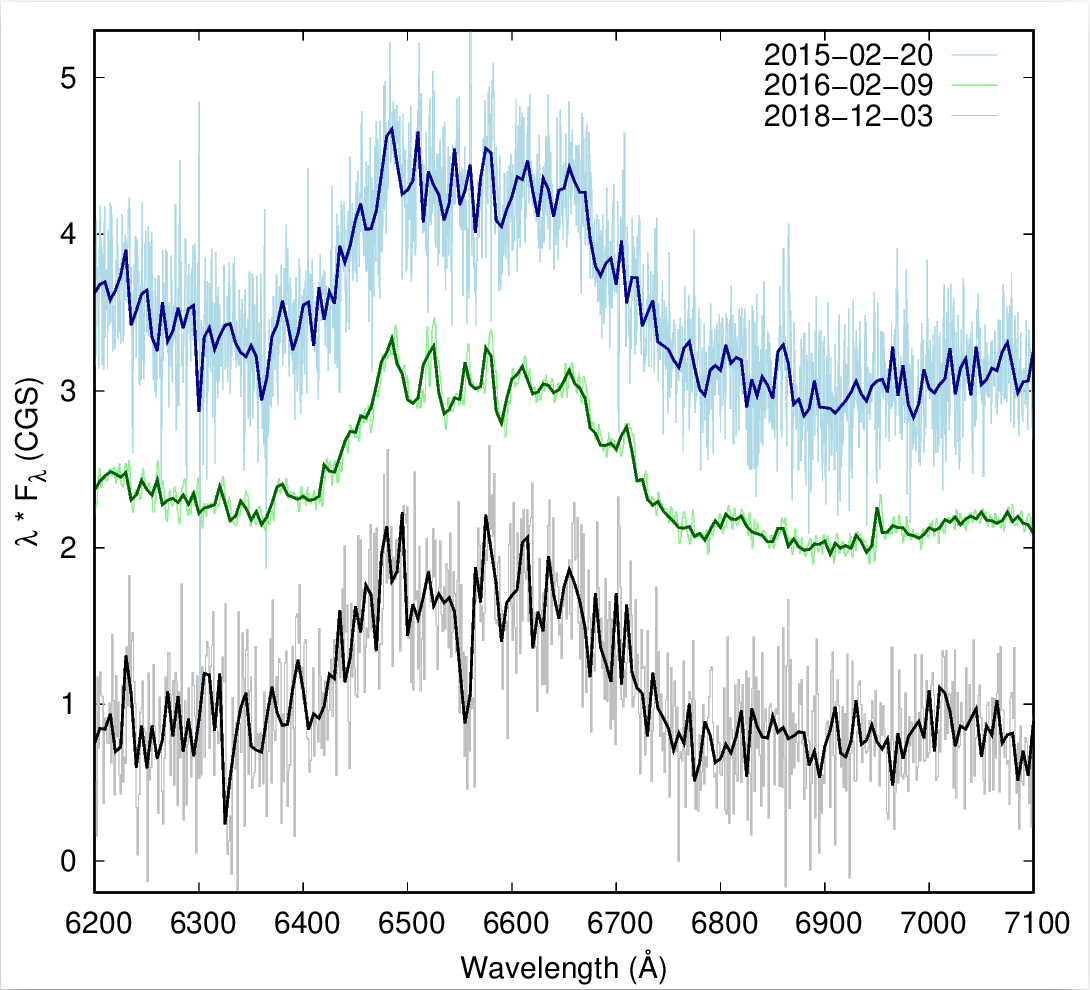}
    \includegraphics[width=0.5\textwidth]{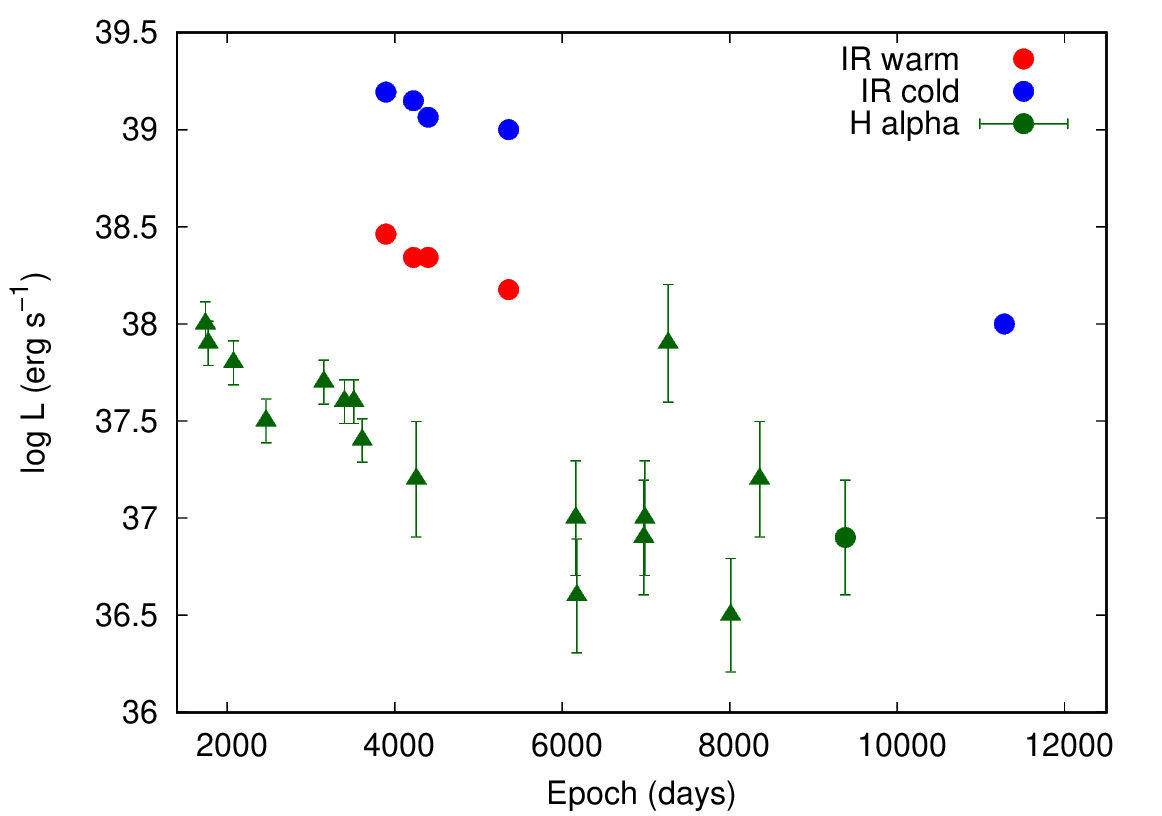}
    \caption{Comparison of late-time H$\alpha$ and IR emission of SN~1993J. {\it Left:} H$\alpha$ emission line profiles in the most recent high-resolution optical spectra of SN~1993J obtained with Keck/LRIS in 2015, 2016 \citep[published by][]{smith17}, and in 2018 (this work); the dates belong to epochs 7999, 8353, and 9381 days, respectively. Spectra are vertically shifted for better comparison; dark solid lines mark the 5 \AA-binned data sets. {\it Right:} Long-term luminosity evolution of warm and cold dust components (Z22 and this work) compared to H$\alpha$ line luminosities \citep[circle: this work; triangles: adopted from][]{smith17} produced dominantly by the part of the SN ejecta crossing and being excited by the reverse shock.}
    \label{fig:93J_lum}
\end{figure*}

Finally, we refer here to the results of \citet{smith17}, who give the long-term evolution of H$\alpha$ luminosities of SN~1993J (also adopting earlier-time data from \citealt{chandra09}). 
They found a slow, continuous decline of log $L_{{\rm H}\alpha}$ from $\sim 10^{38}$ to $\sim 10^{37}$~erg~s$^{-1}$ during days $\sim 2000$--8300 days 
(note, however, that while line profiles seem to be similar, log $L_{{\rm H}\alpha}$ values show large scatter after $\sim 6000$ days).
We highlight here that the breadth of the late-time H$\alpha$ feature ($v \gtrsim$ 5000 km s$^{-1}$) makes this gas in SN 1993J associated with ejecta being excited by the reverse shock; this is different from the strongly interacting SNe IIn where the persistent H$\alpha$ emission ($v \lesssim$ 2500 km s$^{-1}$) is associated with progenitor wind swept up by the forward shock \citep[see a detailed description by, e.g.,][]{smith17,Milisavljevic_17hsn}.
We added one more point measured from the previously unpublished Keck/LRIS optical spectrum obtained at the epoch of day 9381 (see Sec. \ref{sec:obs}), which also seems to fit into this trend (we apply a 30\% uncertainty for the calculated line flux, following the \citealt{smith17} estimation of earlier Keck/LRIS data).
Fig. \ref{fig:93J_lum} shows the comparison of H$\alpha$ line luminosities and IR luminosities calculated from cold and warm dust SED components (the latter values are adopted from Z22). 
The declining trends of all these curves seem similar, which may suggest that longstanding (but continuously weakening) interaction has a role in heating both cold and warm dust components in the environment of SN~1993J.
We discuss the details of possible dust-heating mechanisms in Sec. \ref{sec:disc}.

\subsection{Local background emission at the SN site: Star formation or an IR echo?} \label{sec:anal_echo}

\begin{figure*}
    \centering
    \includegraphics[width=0.95\textwidth]{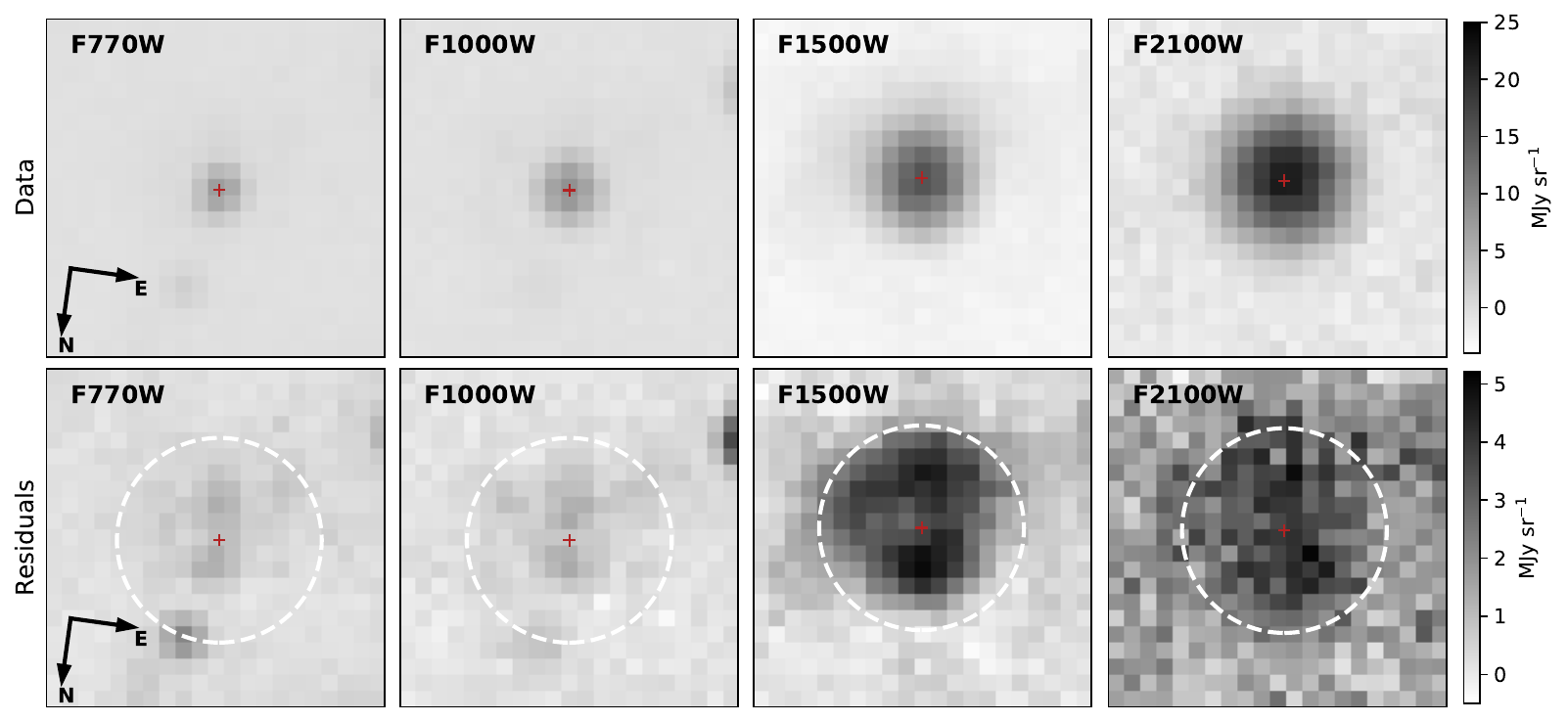}
    \caption{
    Background emission in the {\it JWST}/MIRI images of SN~1993J after removing the bright SN component. {\it Top:} The location of SN~1993J in each of the four filters of MIRI imaging. The red cross indicates indicates the centroid of the SN in each panel obtained from our PSF-fitting photometry. Note that the photometry was performed on individual dithered exposures, but here the pixels around the SN position have been median combined for display purposes.  {\it Bottom:} Background emission around the position of SN~1993J in the four MIRI filters after the removal of the bright SN source during PSF-fitting photometry. In each case, the background level immediately under the SN was determined by fitting a small box (5-pixel width). The fits were repeated with a larger box while holding the background fixed to subtract the SN PSF. As in the top row, the residual data from each dither have been median combined for display purposes, and the red cross indicates the centroid of the SN PSF. The white dashed circle is $0\farcs7$ in radius from this position.
    }
    \label{fig:93J_ring}
\end{figure*}


As described above in Section~\ref{sec:obs}, we also examine the local background environment of SN~1993J based on our detailed photometric PSF-fitting analysis. 
The morphology and integrated fluxes of the residual emission are largely robust for appropriate choices of box sizes during the PSF-fitting procedure. Originally, we first fit the SN with the background flux as a free parameter using a small box (with a 5-pixel width) so that the underlying background is approximately uniform. Then, we fixed the background level to this value and used a larger (21-pixel) box to allow us to integrate the total residual fluxes with the SN removed. 
In Fig.~\ref{fig:93J_ring}, we show the residual emission after removal of the PSF profiles obtained from these fits. The background emission is apparent in all four filters, with the strongest emission at F1500W and F2100W. Comparison with the higher resolution F770W and F1000W suggests the emission may be the result of a combination of multiple, blended point-like sources as well as extended sources. 
To test the robustness of this procedure, we repeated it with small box sizes of 5, 7, and 9 pixels, and large box sizes of 15, 19, and 21 pixels, and find generally consistent results for the integrated residual fluxes of F1500W = $0.11 \pm 0.01$ mJy and F2100W = $0.14 \pm 0.01$ mJy.
As shown in Appendix \ref{sec:appendixB} (Fig. \ref{fig:93J_ring_fit}), using too large a box for the first fit such that the background residuals are nonuniform within the fitting region, results in a spurious ring-like morphology due to oversubtraction of the core of the PSF. 


Based on a simple BB fit to the residual fluxes and upper limits, the nature of the detected background emission seems to be consistent with dust at $\sim 190 \pm 20$~K (see Appendix \ref{sec:appendixB} for more details). The heating source of this dust could be local, ongoing star formation in the vicinity of SN~1993J or alternatively, a thermal IR echo powered by the SN itself. 

We note that \citet{sugerman02} and \citet{liu03} identified optical light echoes (LEs) in {\it HST} images obtained in 2001. These LEs are assumed to scatter from interstellar dust sheets lying $\sim 80$ and $\sim 220$~pc in front of the SN, respectively. The bulk of the emission in {\it JWST}/MIRI F1500W and F2100W images is within $< 0.7$\arcsec, firmly below the radii of the two known optical echoes ($\sim 1.15$\arcsec and $\sim 1.85$--1.95\arcsec). Thus, this emission would correspond to a new echo from dust that, given a delay time of 31~yr, lies largely behind the SN at physical distances of $\sim 5$--13~pc (see, e.g., Eq.~1 of \citealp{dwek08} for the relevant light-echo geometry). 
We note here that similar but more extended IR echoes have been observed around the Galactic SNR Cassiopeia~A (Cas~A; $\sim 150$~K, e.g., \citealt{Krause_2005,dwek08}), which has been found to be the result of a SN~1993J-like Type IIb explosion that occurred $\sim 350$~yr ago \citep{Krause_2008}. \citet{dwek08} found that a short ($\sim 1$~day) burst  of UV radiation with a luminosity of $\sim 1.5 \times 10^{11}$~L$_{\odot}$ from the SN shock breakout was needed to power the observed Cas A echo spectra from dust lying $\sim 50$~pc from the SN. Given the proximity of the dust in the case of SN~1993J, even a less extreme burst could plausibly power the observed emission as an IR echo, but a detailed examination of this possibility is beyond the scope of this work. Additional observations to better constrain the spectrum of the emission or detect variations in its apparent size or brightness would be necessary, for example, to characterize or rule out contributions from ongoing, local star formation to heating the dust.



\section{Discussion}\label{sec:disc}

\begin{figure}
\centering
    \includegraphics[width=\columnwidth]{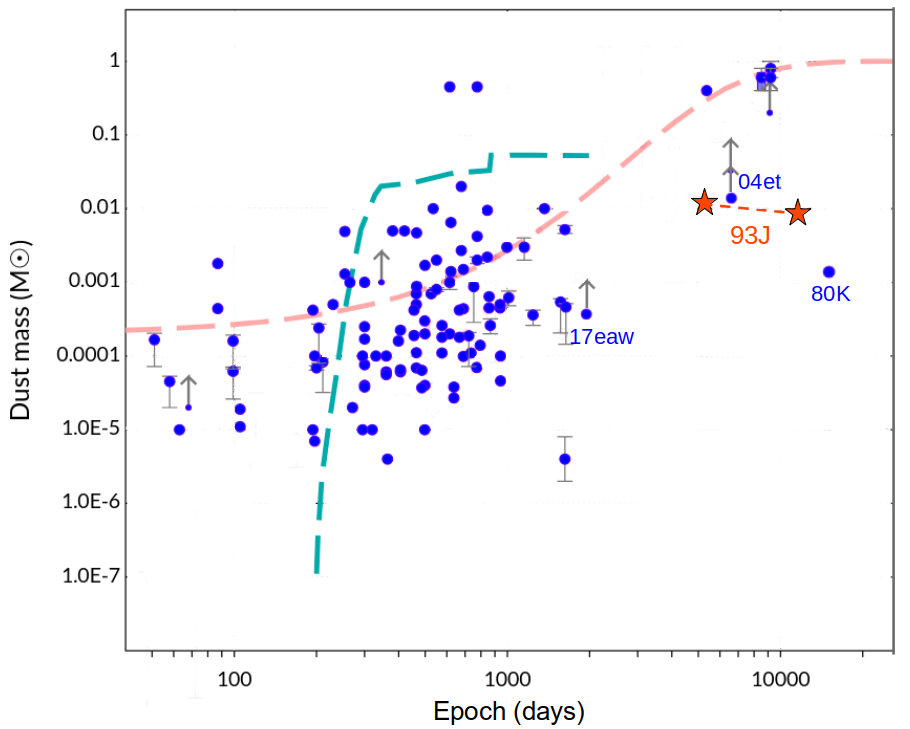}
    \caption{A compilation of literature values for dust mass determinations, based on mid-IR SED analyses, in several individual CCSNe. The orange stars mark the dust masses of SN~1993J determined during this work. Other blue labels mark the cases of SNe IIP 2004et, 2017eaw \citep{Shahbandeh2023}, and SN IIL 1980K \citep{Zsiros_2024}, where the estimated dust masses are also based on the recent analyses of JWST/MIRI SEDs. Dashed lines show examples on rapid \citep[turquoise line,][]{sarangi15} and on steady, continuous \citep[pink line,][]{Wesson15} dust-formation theories.
    The source of the original, interactive diagram is Roger Wesson's homepage (\href{https://nebulousresearch.org/dustmasses/}{https://nebulousresearch.org/dustmasses/}).}
    \label{fig:dustmass}
\end{figure}

As presented above, the model fitting of the current {\it JWST}/MIRI SED is consistent with
the presence of carbonaceous rather than Si-dominated dust, while this was not as clear from the earlier-phase (3893--5359 days) {\it Spitzer} data (Z22).
Assuming thus the presence of pure amorphous carbon dust in SN~1993J and applying updated $\kappa$ values \citep{sarangi22} for both recent {\it JWST} and earlier {\it Spitzer} data, 
we get a similar amount ($\sim 0.01$~\msolar) of dust for now as for $\sim 15$~yr  ago; see Table \ref{tab:dustpar}).
Fig. \ref{fig:dustmass} shows the determined dust masses in SN~1993J compared to other literature values based on mid-IR SED analyses.

Note, however, that the modeled SED parameters determined from the previously obtained {\it Spitzer} data should be handled carefully, since the 8--24~$\mu$m range was not covered then and the spatial resolution of data at those wavelengths is also lower (especially that of the MIPS 24~$\mu$m images).
Moreover, given the lack of any high-quality mid-IR spectra of SN~1993J (or any other SNe of similar type and age), the line-emission contribution to the measured fluxes is also still uncertain.
It seems, however, that the temperature of the cold dust component is significantly lower now than it was $\sim 15$~yr ago ($\sim 120$~K vs. $\sim 165$~K; see Table \ref{tab:dustpar}) --- but note again that the modeling results of the {\it Spitzer} SEDs have strong limitations. 

As shown in Table \ref{tab:dustpar}, we calculate the current luminosity of the cold dust component and obtain $L_\textrm{cold} \approx 1.0 \times 10^{38}$~erg~s$^{-1}$. 
This value, while definitely lower than $\sim 15$~yr ago, is still far above the total luminosity expected from the radioactive decay of $^{44}$Ti $+$ $^{57}$Co $+$ $^{60}$Co (a total of $10^{36}$--$10^{37}$~erg~s$^{-1}$), which is assumed to be the dominant energy source in CCSN ejecta several years or decades after explosion \citep[see, e.g.,][]{tanaka12,Seitenzahl14}.
Thus, an extra energy source is needed to heat the ejecta dust up to the observed temperature. Known longstanding CSM interaction in the environment of SN~1993J is the primary explanation. 
As we show in Fig. \ref{fig:93J_lum}, mid-IR luminosities have a long-term evolution very similar to that measured in H$\alpha$, and the latter is a well-known tracer of the level of shock-CSM interaction. Note, furthermore, that the majority of the interaction luminosity emerges in the form of X-ray and UV photons; as was shown by (for example) \citet{smith17} in the case of some strongly interacting SNe, the X-ray luminosity can achieve a factor of $\sim 10$ higher level than the optical (H$\alpha$) luminosity. Thus, since the latest measured $L_\textrm{{\rm H}$\alpha$}$ values of SN~1993J are in the range $10^{36.5}$--$10^{38}$~erg~s$^{-1}$, the UV/X-ray output from shock-ejecta interaction can be enough to power the observed dust luminosity ($\sim 1.0 \times 10^{38}$~erg~s$^{-1}$) measured by {\it JWST} on day 11,280.
Furthermore, as found by \citet{Dessart_2022} and \cite{Dessart_2023a}, even
modest stellar winds can build up and result in a relatively large
UV flux at late times, generating a constant shock power of about 10$^{40}$ erg s$^{-1}$ in a standard Type II SN.
Depending on the optical depth of the ejecta, back-scattered UV photons may be able to effectively heat not only circumstellar (pre-existing) dust but also newly-formed grains located in the inner ejecta, as we discussed in more detail by \citet{Shahbandeh2023}. 
Nevertheless, note that collecting direct observational UV evidence for the latter option seems to be quite challenging at such late epochs.

Finally, taking a closer look at the Galactic SNR Cas~A could be  helpful for understanding the dust origin and heating mechanisms in the case of SN~1993J (and in other Type IIb explosions).
Studying the dust content of Cas~A was the topic of numerous studies \citep[e.g.,][]{dwek08,dunne09,barlow10,arendt14,bevan17,delooze17,Priestley_2019,Priestley_2022,Kirchschlager_2023,Kirchschlager_2024} and the object has also been a target of recent {\it JWST} observations \citep{Milisavljevic_2024}. While we see many exciting details in the structure and ongoing interaction processes of this nearby SNR, there are efforts to understand its past and the fate of SNe~IIb in general. SN~1993J, already in the transitional phase between an SN and an SNR, is an ideal object for follow-up studies from this perspective as well.

The cited studies present detailed investigations of the physical composition of the dust in both the ejecta and the ambient swept-up medium of Cas~A. A general conclusion of these investigations is that the mid-IR component of Cas~A's spectrum basically emerges from the post-shocked ejecta region, which contains a few $\times 10^{-3}$~\msolar\ of dust.
This value is in good agreement with the one we found during the analysis of our current {\it JWST} data of SN~1993J. It may indicate that the dust we see now in SN~1993J can also be in the post-shocked ejecta.
Furthermore, as these studies suggest, a much larger amount ($\sim 0.1$--1.0~\msolar) of very cold ($< 100$~K) dust is located within the unshocked ejecta of Cas~A. Following the arguments we described above, it could be possible that we do see part of the heated unshocked ejecta dust in the mid-IR data of SN~1993J now, and this dust can cool well below 100~K after several decades (as the intensity of the CSM interaction decreases, suggested by the current trend seen in Fig. \ref{fig:93J_lum}).

In the case of ejecta dust, the effect of grain-destruction processes caused by the reverse shock should be taken into account as well. This topic has been actively studied in the case of Cas~A \citep[e.g.,][]{Micelotta_2016,delooze17,Priestley_2022,Kirchschlager_2023,Kirchschlager_2024}, especially because theoretical studies predict a small average grain size ($a<0.01$~$\mu$m) in SN~IIb ejecta \citep{nozawa10,Biscaro_2014,Biscaro_2016}. While all these studies predict a very low survival rate of grains, the estimated rate values depend on various parameters (grain-size distribution, density contrast between the dust clumps and the ambient medium,  magnetic field strength, etc.) and seem to increase after some time; see the most recent calculations by \citet{Kirchschlager_2023,Kirchschlager_2024}. However, as can be inferred from these studies, note that the reverse shock may have hit only the outermost part of the ejecta of SN~1993J by now; thus, the dust content in the inner part should not be affected yet. This is in agreement with our finding of similar dust masses in SN~1993J at $\sim 15$ and $\sim 30$~yr after explosion.

We also remark that the dust composition used to describe Cas~A's IR-submm spectra is mainly based on Al$_2$O$_3$ and various types of magnesium silicates \citep[e.g.,][]{arendt14}, but dust-synthesis models also allow for the presence of larger amounts of carbonaceous grains in SN~IIb explosions \citep{nozawa10,Biscaro_2014,Biscaro_2016}.
Note, however, that lack of the characteristic silicate features does not necessarily mean that we do really see carbonaceous dust. It could also be very large silicate grains or high optical depths that quash the signs of silicate emission.
A future {\it JWST} spectrum would be the necessary further step in revealing the true composition of dust in SN~1993J.


\section{Conclusions}\label{sec:conc}

To summarize, analysis of recent {\it JSWT}/MIRI photometric data of SN~1993J suggests that it still contains a significant amount of dust $\sim 30$~yr after explosion. Comparing the current results to those of the analysis of earlier {\it Spitzer} data, 
we see a similar amount ($\sim 0.01$~\msolar) of carbonaceous dust to that detected $\sim 15$--20~yr ago, but at a lower temperature ($\sim 120$~K vs. $\sim 165$~K).

There are still open questions regarding whether we see the {\it same} dust, and whether this dust has formed post-explosion (and  is located in the inner or the post-shocked ejecta) or is pre-existing.
As shown by Z22, this amount of dust can be entirely in the inner (unshocked) ejecta of SN~1993J, supported also by the red–blue line-profile asymmetries of late-time optical spectra. In this case, it is possible that we see the same dust now (since we expect no grain destruction in the unshocked region).
This dust, if located in the inner ejecta, has  presumably cooled over the years. This is also in agreement with the continuously decreasing intensity of the CSM interaction, which is assumed to be the main heating mechanism even for ejecta dust (the presence of still ongoing CSM interaction is revealed not just in the form of the long-lived H$\alpha$ emission-line profile but also as a hot component in the recent {\it JWST}/MIRI SED of SN~1993J).

As another possibility  (also shown by Z22), part or all of the observed dust (either pre-existing or newly formed) can be in the post-shocked regions of the SN. We see such dust in the mid-IR in the $\sim 350$~yr-old remnant of Cas~A, in an amount similar to what was found in SN~1993J at a much younger age. In general, we barely have any information on the dust evolution in the SN-SNR transitional phase; however, one can assume (for example) continuous destruction and condensation of grains. 
Note, furthermore, that the dust amount ($\sim 0.01$~ \msolar) found in SN~1993J based on {\it JWST}/MIRI data is comparable to dust masses seen in strongly interacting Type IIn SNe \citep[e.g.,][]{fox11,fox13,Fox_2020}. Since we assume much lower mass-loss rates in SN~IIb progenitors than in SN~IIn ones, it may be an argument against entirely pre-existing dust in SN~1993J. Note, however, that most of the known SN~IIn IR dust masses were determined from short-wavelength {\it Spitzer} data --- upcoming {\it JWST} observations of SNe~IIn and uniform dust-modeling methods (especially on grain properties) are expected to achieve further progress in this field.

Moreover, beyond the local dust assumed in the close environment of SN~1993J, we may have identified signs of a potential IR echo (found as residual background emission in {\it JWST}/MIRI images after PSF subtraction) --- that is, radiation of a more-distant dust shell heated by the SN shock-breakout luminosity.
Nevertheless, to get a full picture of the dusty SN~1993J, it would be useful to obtain a complete near-IR and mid-IR spectrum at the same SN phase. 
Such a dataset would allow us to take into account further important aspects such as the effects of  atomic and molecular line emission on the IR spectrum \citep[see][for the case of Cas~A]{Milisavljevic_2024}, or how the progenitor's binary companion affects the CSM geometry and the dust formation/heating processes \citep[e.g.,][]{fox14,Kochanek_2017}.
All these arguments, along with the results presented above, should make SN~1993J a promising spectroscopic target in future {\it JWST} cycles and the subject of further detailed studies.

\section*{Data availability}

Data are available at the 
\href{https://mast.stsci.edu/search/ui/#/jwst/results?resolve=true&target=sn1993j&data_types=image,measurements&instruments=MIRI&pi_surname=Fox&radius=3&radius_units=arcminutes&useStore=false&search_key=e44bbb62f2eeb}{Barbara A. Mikulski Archive for Space Telescopes} (MAST).

\begin{acknowledgements}

We thank our anonymous referee for valuable comments.
This work is based on observations made with the NASA/ESA/CSA {\it James Webb Space Telescope}. The data were obtained from the Mikulski Archive for Space Telescopes at the Space Telescope Science Institute, which is operated by the Association of Universities for Research in Astronomy, Inc., under NASA contract NAS 5-03127 for {\it JWST}. These observations are associated with program GO-3921.

This project has been supported by the NKFIH OTKA FK-134432 grant of the National Research, Development and Innovation (NRDI) Office of Hungary. S.Z. is supported by the {\'U}NKP-23-4-SZTE-574 New National Excellence Program of the Ministry for Culture and Innovation from the source of the NRDI Fund, Hungary. I.D.L. has received funding from the European Research Council (ERC) under the European Union's Horizon 2020 research and innovation programme DustOrigin (ERC-2019-StG-851622) and the Belgian Science Policy Office (BELSPO) through the PRODEX project ``JWST/MIRI Science exploitation'' (C4000142239).
A.V.F. is grateful for financial support from the Christopher R. Redlich Fund and many other donors.
C.A. acknowledges support by NASA JWST grants GO-02114, GO-02122, GO-03726, GO-04436, and GO-04522.

Some of the data presented herein were obtained at the W. M. Keck
Observatory, which is operated as a scientific partnership among the
California Institute of Technology, the University of California, and
NASA; the observatory was made possible by the generous financial
support of the W. M. Keck Foundation.

\end{acknowledgements}

%
%
%
\bibliographystyle{aa}
\bibliography{main} 

\begin{thebibliography}{102}
\expandafter\ifx\csname natexlab\endcsname\relax\def\natexlab#1{#1}\fi

\bibitem[{{Aldering} {et~al.}(1994){Aldering}, {Humphreys}, \&
  {Richmond}}]{aldering94}
{Aldering}, G., {Humphreys}, R.~M., \& {Richmond}, M. 1994, \aj, 107, 662

\bibitem[{{Arendt} {et~al.}(2016){Arendt}, {Dwek}, {Bouchet}, {Danziger},
  {Frank}, {Gehrz}, {Park}, \& {Woodward}}]{arendt16}
{Arendt}, R.~G., {Dwek}, E., {Bouchet}, P., {et~al.} 2016, \aj, 151, 62

\bibitem[{{Arendt} {et~al.}(2020){Arendt}, {Dwek}, {Bouchet}, {John Danziger},
  {Gehrz}, {Park}, \& {Woodward}}]{arendt20}
{Arendt}, R.~G., {Dwek}, E., {Bouchet}, P., {et~al.} 2020, \apj, 890, 2

\bibitem[{{Arendt} {et~al.}(2014){Arendt}, {Dwek}, {Kober}, {Rho}, \&
  {Hwang}}]{arendt14}
{Arendt}, R.~G., {Dwek}, E., {Kober}, G., {Rho}, J., \& {Hwang}, U. 2014, \apj,
  786, 55

\bibitem[{{Baer-Way} {et~al.}(2024){Baer-Way}, {DeGraw}, {Zheng}, {Van Dyk},
  {Filippenko}, {Fox}, {Brink}, {Kelly}, {Smith}, {Vasylyev}, {de Jaeger},
  {Zhang}, {Stegman}, {Ross}, \& {Yunus}}]{Baer-Way_2024}
{Baer-Way}, R., {DeGraw}, A., {Zheng}, W., {et~al.} 2024, \apj, 964, 172

\bibitem[{{Barlow} {et~al.}(2010){Barlow}, {Krause}, {Swinyard}, {Sibthorpe},
  {Besel}, {Wesson}, {Ivison}, {Dunne}, {Gear}, {Gomez}, {Hargrave}, {Henning},
  {Leeks}, {Lim}, {Olofsson}, \& {Polehampton}}]{barlow10}
{Barlow}, M.~J., {Krause}, O., {Swinyard}, B.~M., {et~al.} 2010, \aap, 518,
  L138

\bibitem[{{Bartel} {et~al.}(2002){Bartel}, {Bietenholz}, {Rupen}, {Beasley},
  {Graham}, {Altunin}, {Venturi}, {Umana}, {Cannon}, \& {Conway}}]{bartel02}
{Bartel}, N., {Bietenholz}, M.~F., {Rupen}, M.~P., {et~al.} 2002, \apj, 581,
  404

\bibitem[{{Bevan} {et~al.}(2017){Bevan}, {Barlow}, \&
  {Milisavljevic}}]{bevan17}
{Bevan}, A., {Barlow}, M.~J., \& {Milisavljevic}, D. 2017, \mnras, 465, 4044

\bibitem[{{Bietenholz} {et~al.}(2001){Bietenholz}, {Bartel}, \&
  {Rupen}}]{bietenholz01}
{Bietenholz}, M.~F., {Bartel}, N., \& {Rupen}, M.~P. 2001, \apj, 557, 770

\bibitem[{{Bietenholz} {et~al.}(2003){Bietenholz}, {Bartel}, \&
  {Rupen}}]{bietenholz03}
{Bietenholz}, M.~F., {Bartel}, N., \& {Rupen}, M.~P. 2003, \apj, 597, 374

\bibitem[{{Biscaro} \& {Cherchneff}(2014)}]{Biscaro_2014}
{Biscaro}, C. \& {Cherchneff}, I. 2014, \aap, 564, A25

\bibitem[{{Biscaro} \& {Cherchneff}(2016)}]{Biscaro_2016}
{Biscaro}, C. \& {Cherchneff}, I. 2016, \aap, 589, A132

\bibitem[{{Bouchet} {et~al.}(2006){Bouchet}, {Dwek}, {Danziger}, {Arendt}, {De
  Buizer}, {Park}, {Suntzeff}, {Kirshner}, \& {Challis}}]{bouchet06}
{Bouchet}, P., {Dwek}, E., {Danziger}, J., {et~al.} 2006, \apj, 650, 212

\bibitem[{{Bouchet} {et~al.}(2015){Bouchet}, {Garc{\'\i}a-Mar{\'\i}n},
  {Lagage}, {Amiaux}, {Augu{\'e}res}, {Bauwens}, {Blommaert}, {Chen}, {Detre},
  {Dicken}, {Dubreuil}, {Galdemard}, {Gastaud}, {Glasse}, {Gordon}, {Gougnaud},
  {Guillard}, {Justtanont}, {Krause}, {Leboeuf}, {Longval}, {Martin}, {Mazy},
  {Moreau}, {Olofsson}, {Ray}, {Rees}, {Renotte}, {Ressler}, {Ronayette},
  {Salasca}, {Scheithauer}, {Sykes}, {Thelen}, {Wells}, {Wright}, \&
  {Wright}}]{Bouchet_2015}
{Bouchet}, P., {Garc{\'\i}a-Mar{\'\i}n}, M., {Lagage}, P.~O., {et~al.} 2015,
  \pasp, 127, 612

\bibitem[{{Brooker} {et~al.}(2022){Brooker}, {Stangl}, {Mauney}, \&
  {Fryer}}]{Brooker_2022}
{Brooker}, E.~S., {Stangl}, S.~M., {Mauney}, C.~M., \& {Fryer}, C.~L. 2022,
  \apj, 931, 85

\bibitem[{{Chandra} {et~al.}(2009){Chandra}, {Dwarkadas}, {Ray}, {Immler}, \&
  {Pooley}}]{chandra09}
{Chandra}, P., {Dwarkadas}, V.~V., {Ray}, A., {Immler}, S., \& {Pooley}, D.
  2009, \apj, 699, 388

\bibitem[{{Cohen} {et~al.}(1995){Cohen}, {Darling}, \& {Porter}}]{cohen95}
{Cohen}, J.~G., {Darling}, J., \& {Porter}, A. 1995, \aj, 110, 308

\bibitem[{{De Looze} {et~al.}(2019){De Looze}, {Barlow}, {Bandiera}, {Bevan},
  {Bietenholz}, {Chawner}, {Gomez}, {Matsuura}, {Priestley}, \&
  {Wesson}}]{delooze19}
{De Looze}, I., {Barlow}, M.~J., {Bandiera}, R., {et~al.} 2019, \mnras, 488,
  164

\bibitem[{{De Looze} {et~al.}(2017){De Looze}, {Barlow}, {Swinyard}, {Rho},
  {Gomez}, {Matsuura}, \& {Wesson}}]{delooze17}
{De Looze}, I., {Barlow}, M.~J., {Swinyard}, B.~M., {et~al.} 2017, \mnras, 465,
  3309

\bibitem[{{Dessart} {et~al.}(2023){Dessart}, {Guti{\'e}rrez}, {Kuncarayakti},
  {Fox}, \& {Filippenko}}]{Dessart_2023a}
{Dessart}, L., {Guti{\'e}rrez}, C.~P., {Kuncarayakti}, H., {Fox}, O.~D., \&
  {Filippenko}, A.~V. 2023, \aap, 675, A33

\bibitem[{{Dessart} \& {Hillier}(2022)}]{Dessart_2022}
{Dessart}, L. \& {Hillier}, D.~J. 2022, \aap, 660, L9

\bibitem[{{Draine} \& {Li}(2007)}]{Draine_2007}
{Draine}, B.~T. \& {Li}, A. 2007, \apj, 657, 810

\bibitem[{{Dunne} {et~al.}(2009){Dunne}, {Maddox}, {Ivison}, {Rudnick},
  {Delaney}, {Matthews}, {Crowe}, {Gomez}, {Eales}, \& {Dye}}]{dunne09}
{Dunne}, L., {Maddox}, S.~J., {Ivison}, R.~J., {et~al.} 2009, \mnras, 394, 1307

\bibitem[{{Dwek} \& {Arendt}(2008)}]{dwek08}
{Dwek}, E. \& {Arendt}, R.~G. 2008, \apj, 685, 976

\bibitem[{{Dwek} {et~al.}(2010){Dwek}, {Arendt}, {Bouchet}, {Burrows},
  {Challis}, {Danziger}, {De Buizer}, {Gehrz}, {Park}, {Polomski}, {Slavin}, \&
  {Woodward}}]{dwek10}
{Dwek}, E., {Arendt}, R.~G., {Bouchet}, P., {et~al.} 2010, \apj, 722, 425

\bibitem[{Dwek {et~al.}(2007)Dwek, Galliano, \& Jones}]{dwek07}
Dwek, E., Galliano, F., \& Jones, A.~P. 2007, The Astrophysical Journal, 662,
  927

\bibitem[{{Dwek} {et~al.}(2019){Dwek}, {Sarangi}, \& {Arendt}}]{Dwek_2019}
{Dwek}, E., {Sarangi}, A., \& {Arendt}, R.~G. 2019, \apjl, 871, L33

\bibitem[{{Dwek} {et~al.}(2021){Dwek}, {Sarangi}, {Arendt}, {Kallman},
  {Kazanas}, \& {Fox}}]{Dwek_2021}
{Dwek}, E., {Sarangi}, A., {Arendt}, R.~G., {et~al.} 2021, \apj, 917, 84

\bibitem[{{Fabbri} {et~al.}(2011){Fabbri}, {Otsuka}, {Barlow}, {Gallagher},
  {Wesson}, {Sugerman}, {Clayton}, {Meixner}, {Andrews}, {Welch}, \&
  {Ercolano}}]{fabbri11}
{Fabbri}, J., {Otsuka}, M., {Barlow}, M.~J., {et~al.} 2011, \mnras, 418, 1285

\bibitem[{{Filippenko}(1982)}]{Filippenko_1982}
{Filippenko}, A.~V. 1982, \pasp, 94, 715

\bibitem[{{Filippenko}(1988)}]{filippenko88}
{Filippenko}, A.~V. 1988, \aj, 96, 1941

\bibitem[{{Filippenko}(1997)}]{Filippenko_97}
{Filippenko}, A.~V. 1997, \araa, 35, 309

\bibitem[{{Filippenko} {et~al.}(1993){Filippenko}, {Matheson}, \&
  {Ho}}]{filippenko93}
{Filippenko}, A.~V., {Matheson}, T., \& {Ho}, L.~C. 1993, \apjl, 415, L103

\bibitem[{{Fox} {et~al.}(2014){Fox}, {Azalee Bostroem}, {Van Dyk},
  {Filippenko}, {Fransson}, {Matheson}, {Cenko}, {Chandra}, {Dwarkadas}, {Li},
  {Parker}, \& {Smith}}]{fox14}
{Fox}, O.~D., {Azalee Bostroem}, K., {Van Dyk}, S.~D., {et~al.} 2014, \apj,
  790, 17

\bibitem[{{Fox} {et~al.}(2010){Fox}, {Chevalier}, {Dwek}, {Skrutskie},
  {Sugerman}, \& {Leisenring}}]{fox10}
{Fox}, O.~D., {Chevalier}, R.~A., {Dwek}, E., {et~al.} 2010, \apj, 725, 1768

\bibitem[{{Fox} {et~al.}(2011){Fox}, {Chevalier}, {Skrutskie}, {Soderberg},
  {Filippenko}, {Ganeshalingam}, {Silverman}, {Smith}, \& {Steele}}]{fox11}
{Fox}, O.~D., {Chevalier}, R.~A., {Skrutskie}, M.~F., {et~al.} 2011, \apj, 741,
  7

\bibitem[{{Fox} {et~al.}(2013){Fox}, {Filippenko}, {Skrutskie}, {Silverman},
  {Ganeshalingam}, {Cenko}, \& {Clubb}}]{fox13}
{Fox}, O.~D., {Filippenko}, A.~V., {Skrutskie}, M.~F., {et~al.} 2013, \aj, 146,
  2

\bibitem[{{Fox} {et~al.}(2020){Fox}, {Fransson}, {Smith}, {Andrews}, {Azalee
  Bostroem}, {Brink}, {Bradley Cenko}, {Clayton}, {Filippenko}, {Fong},
  {Gallagher}, {Kelly}, {Kilpatrick}, {Mauerhan}, {Miller}, {Montiel},
  {Stritzinger}, {Szalai}, \& {Van Dyk}}]{Fox_2020}
{Fox}, O.~D., {Fransson}, C., {Smith}, N., {et~al.} 2020, \mnras, 498, 517

\bibitem[{{Fransson} \& {Kozma}(2002)}]{fransson02}
{Fransson}, C. \& {Kozma}, C. 2002, \nar, 46, 487

\bibitem[{{Freedman} {et~al.}(2001){Freedman}, {Madore}, {Gibson}, {Ferrarese},
  {Kelson}, {Sakai}, {Mould}, {Kennicutt}, {Ford}, {Graham}, {Huchra},
  {Hughes}, {Illingworth}, {Macri}, \& {Stetson}}]{freedman01}
{Freedman}, W.~L., {Madore}, B.~F., {Gibson}, B.~K., {et~al.} 2001, \apj, 553,
  47

\bibitem[{{Gall} {et~al.}(2014){Gall}, {Hjorth}, {Watson}, {Dwek}, {Maund},
  {Fox}, {Leloudas}, {Malesani}, \& {Day-Jones}}]{gall14}
{Gall}, C., {Hjorth}, J., {Watson}, D., {et~al.} 2014, \nat, 511, 326

\bibitem[{{Gomez} {et~al.}(2012){Gomez}, {Krause}, {Barlow}, {Swinyard},
  {Owen}, {Clark}, {Matsuura}, {Gomez}, {Rho}, {Besel}, {Bouwman}, {Gear},
  {Henning}, {Ivison}, {Polehampton}, \& {Sibthorpe}}]{gomez12}
{Gomez}, H.~L., {Krause}, O., {Barlow}, M.~J., {et~al.} 2012, The Astrophysical
  Journal, 760, 96

\bibitem[{{Hildebrand}(1983)}]{hildebrand83}
{Hildebrand}, R.~H. 1983, \qjras, 24, 267

\bibitem[{{Indebetouw} {et~al.}(2014){Indebetouw}, {Matsuura}, {Dwek},
  {Zanardo}, {Barlow}, {Baes}, {Bouchet}, {Burrows}, {Chevalier}, {Clayton},
  {Fransson}, {Gaensler}, {Kirshner}, {Laki{\'c}evi{\'c}}, {Long}, {Lundqvist},
  {Mart{\'\i}-Vidal}, {Marcaide}, {McCray}, {Meixner}, {Ng}, {Park},
  {Sonneborn}, {Staveley-Smith}, {Vlahakis}, \& {van Loon}}]{indebetouw14}
{Indebetouw}, R., {Matsuura}, M., {Dwek}, E., {et~al.} 2014, ApJL, 782, L2

\bibitem[{{Jones} {et~al.}(2023){Jones}, {Kavanagh}, {Barlow}, {Temim},
  {Fransson}, {Larsson}, {Blommaert}, {Meixner}, {Lau}, {Sargent}, {Bouchet},
  {Hjorth}, {Wright}, {Coulais}, {Fox}, {Gastaud}, {Glasse}, {Habel},
  {Hirschauer}, {Jaspers}, {Krause}, {Lenki{\'c}}, {Nayak}, {Rest}, {Tikkanen},
  {Wesson}, {Colina}, {van Dishoeck}, {G{\"u}del}, {Henning}, {Lagage},
  {{\"O}stlin}, {Ray}, \& {Vandenbussche}}]{Jones_2023}
{Jones}, O.~C., {Kavanagh}, P.~J., {Barlow}, M.~J., {et~al.} 2023, \apj, 958,
  95

\bibitem[{{Kirchschlager} {et~al.}(2024){Kirchschlager}, {Sartorio}, {De
  Looze}, {Barlow}, {Schmidt}, \& {Priestley}}]{Kirchschlager_2024}
{Kirchschlager}, F., {Sartorio}, N.~S., {De Looze}, I., {et~al.} 2024, \mnras,
  528, 5364

\bibitem[{{Kirchschlager} {et~al.}(2023){Kirchschlager}, {Schmidt}, {Barlow},
  {De Looze}, \& {Sartorio}}]{Kirchschlager_2023}
{Kirchschlager}, F., {Schmidt}, F.~D., {Barlow}, M.~J., {De Looze}, I., \&
  {Sartorio}, N.~S. 2023, \mnras, 520, 5042

\bibitem[{{Kochanek}(2017)}]{Kochanek_2017}
{Kochanek}, C.~S. 2017, \mnras, 471, 3283

\bibitem[{{Kotak} {et~al.}(2009){Kotak}, {Meikle}, {Farrah}, {Gerardy},
  {Foley}, {Van Dyk}, {Fransson}, {Lundqvist}, {Sollerman}, {Fesen},
  {Filippenko}, {Mattila}, {Silverman}, {Andersen}, {H{\"o}flich}, {Pozzo}, \&
  {Wheeler}}]{kotak09}
{Kotak}, R., {Meikle}, W.~P.~S., {Farrah}, D., {et~al.} 2009, \apj, 704, 306

\bibitem[{{Krause} {et~al.}(2008){Krause}, {Birkmann}, {Usuda}, {Hattori},
  {Goto}, {Rieke}, \& {Misselt}}]{Krause_2008}
{Krause}, O., {Birkmann}, S.~M., {Usuda}, T., {et~al.} 2008, Science, 320, 1195

\bibitem[{{Krause} {et~al.}(2005){Krause}, {Rieke}, {Birkmann}, {Le Floc'h},
  {Gordon}, {Egami}, {Bieging}, {Hughes}, {Young}, {Hinz}, {Quanz}, \&
  {Hines}}]{Krause_2005}
{Krause}, O., {Rieke}, G.~H., {Birkmann}, S.~M., {et~al.} 2005, Science, 308,
  1604

\bibitem[{{Kundu} {et~al.}(2019){Kundu}, {Lundqvist}, {Sorokina},
  {P{\'e}rez-Torres}, {Blinnikov}, {O'Connor}, {Ergon}, {Chandra}, \&
  {Das}}]{kundu19}
{Kundu}, E., {Lundqvist}, P., {Sorokina}, E., {et~al.} 2019, \apj, 875, 17

\bibitem[{{Larsson} {et~al.}(2023){Larsson}, {Fransson}, {Sargent}, {Jones},
  {Barlow}, {Bouchet}, {Meixner}, {Blommaert}, {Coulais}, {Fox}, {Gastaud},
  {Glasse}, {Habel}, {Hirschauer}, {Hjorth}, {Jaspers}, {Kavanagh}, {Krause},
  {Lau}, {Lenki{\'c}}, {Nayak}, {Rest}, {Temim}, {Tikkanen}, {Wesson}, \&
  {Wright}}]{Larsson_2023}
{Larsson}, J., {Fransson}, C., {Sargent}, B., {et~al.} 2023, \apjl, 949, L27

\bibitem[{{Lewis} {et~al.}(1994){Lewis}, {Walton}, {Meikle}, {Martin},
  {Cumming}, {Catchpole}, {Arevalo}, {Argyle}, {Benn}, {Bunclark}, {Castaneda},
  {Centurion}, {Clegg}, {Delgado}, {Dhillon}, {Goudfrooij}, {Harlaftis},
  {Hassall}, {Helmer}, {Hill}, {Jones}, {King}, {Lazaro}, {Lucey}, {Martin},
  {Miller}, {Morrison}, {Penny}, {Perez}, {Read}, {Rudd}, {Rutten}, {Sharples},
  {Unger}, \& {Vilchez}}]{lewis94}
{Lewis}, J.~R., {Walton}, N.~A., {Meikle}, W.~P.~S., {et~al.} 1994, \mnras,
  266, L27

\bibitem[{{Liu} {et~al.}(2003){Liu}, {Bregman}, \& {Seitzer}}]{liu03}
{Liu}, J.-F., {Bregman}, J.~N., \& {Seitzer}, P. 2003, \apj, 582, 919

\bibitem[{{Mart{\'\i}-Vidal} {et~al.}(2011){Mart{\'\i}-Vidal}, {Marcaide},
  {Alberdi}, {Guirado}, {P{\'e}rez-Torres}, \& {Ros}}]{martividal11}
{Mart{\'\i}-Vidal}, I., {Marcaide}, J.~M., {Alberdi}, A., {et~al.} 2011, \aap,
  526, A142

\bibitem[{{Matheson} {et~al.}(2000){Matheson}, {Filippenko}, {Ho}, {Barth}, \&
  {Leonard}}]{matheson00}
{Matheson}, T., {Filippenko}, A.~V., {Ho}, L.~C., {Barth}, A.~J., \& {Leonard},
  D.~C. 2000, \aj, 120, 1499

\bibitem[{{Matsuura} {et~al.}(2019){Matsuura}, {De Buizer}, {Arendt}, {Dwek},
  {Barlow}, {Bevan}, {Cigan}, {Gomez}, {Rho}, {Wesson}, {Bouchet}, {Danziger},
  \& {Meixner}}]{matsuura19}
{Matsuura}, M., {De Buizer}, J.~M., {Arendt}, R.~G., {et~al.} 2019, \mnras,
  482, 1715

\bibitem[{{Matsuura} {et~al.}(2011){Matsuura}, {Dwek}, {Meixner}, {Otsuka},
  {Babler}, {Barlow}, {Roman-Duval}, {Engelbracht}, {Sandstrom},
  {Laki{\'c}evi{\'c}}, {van Loon}, {Sonneborn}, {Clayton}, {Long}, {Lundqvist},
  {Nozawa}, {Gordon}, {Hony}, {Panuzzo}, {Okumura}, {Misselt}, {Montiel}, \&
  {Sauvage}}]{matsuura11}
{Matsuura}, M., {Dwek}, E., {Meixner}, M., {et~al.} 2011, Science, 333, 1258

\bibitem[{{Matthews} {et~al.}(2002){Matthews}, {Neugebauer}, {Armus}, \&
  {Soifer}}]{matthews02}
{Matthews}, K., {Neugebauer}, G., {Armus}, L., \& {Soifer}, B.~T. 2002, \aj,
  123, 753

\bibitem[{{Maund} \& {Smartt}(2009)}]{maund09}
{Maund}, J.~R. \& {Smartt}, S.~J. 2009, Science, 324, 486

\bibitem[{{Maund} {et~al.}(2004){Maund}, {Smartt}, {Kudritzki},
  {Podsiadlowski}, \& {Gilmore}}]{maund04}
{Maund}, J.~R., {Smartt}, S.~J., {Kudritzki}, R.~P., {Podsiadlowski}, P., \&
  {Gilmore}, G.~F. 2004, \nat, 427, 129

\bibitem[{{Meikle} {et~al.}(2011){Meikle}, {Kotak}, {Farrah}, {Mattila}, {Van
  Dyk}, {Andersen}, {Fesen}, {Filippenko}, {Foley}, {Fransson}, {Gerardy},
  {H{\"o}flich}, {Lundqvist}, {Pozzo}, {Sollerman}, \& {Wheeler}}]{meikle11}
{Meikle}, W.~P.~S., {Kotak}, R., {Farrah}, D., {et~al.} 2011, \apj, 732, 109

\bibitem[{{Micelotta} {et~al.}(2016){Micelotta}, {Dwek}, \&
  {Slavin}}]{Micelotta_2016}
{Micelotta}, E.~R., {Dwek}, E., \& {Slavin}, J.~D. 2016, \aap, 590, A65

\bibitem[{{Milisavljevic} \& {Fesen}(2017)}]{Milisavljevic_17hsn}
{Milisavljevic}, D. \& {Fesen}, R.~A. 2017, in Handbook of Supernovae, ed.
  A.~W. {Alsabti} \& P.~{Murdin}, 2211

\bibitem[{{Milisavljevic} {et~al.}(2012){Milisavljevic}, {Fesen}, {Chevalier},
  {Kirshner}, {Challis}, \& {Turatto}}]{milisavljevic12}
{Milisavljevic}, D., {Fesen}, R.~A., {Chevalier}, R.~A., {et~al.} 2012, \apj,
  751, 25

\bibitem[{{Milisavljevic} {et~al.}(2024){Milisavljevic}, {Temim}, {De Looze},
  {Dickinson}, {Laming}, {Fesen}, {Raymond}, {Arendt}, {Vink}, {Posselt},
  {Pavlov}, {Fox}, {Pinarski}, {Subrayan}, {Schmidt}, {Blair}, {Rest},
  {Patnaude}, {Koo}, {Rho}, {Orlando}, {Janka}, {Andrews}, {Barlow}, {Burrows},
  {Chevalier}, {Clayton}, {Fransson}, {Fryer}, {Gomez}, {Kirchschlager}, {Lee},
  {Matsuura}, {Niculescu-Duvaz}, {Pierel}, {Plucinsky}, {Priestley}, {Ravi},
  {Sartorio}, {Schmidt}, {Shahbandeh}, {Slane}, {Smith}, {Sravan}, {Weil},
  {Wesson}, \& {Wheeler}}]{Milisavljevic_2024}
{Milisavljevic}, D., {Temim}, T., {De Looze}, I., {et~al.} 2024, \apjl, 965,
  L27

\bibitem[{{Nomoto} {et~al.}(1993){Nomoto}, {Suzuki}, {Shigeyama}, {Kumagai},
  {Yamaoka}, \& {Saio}}]{nomoto93}
{Nomoto}, K., {Suzuki}, T., {Shigeyama}, T., {et~al.} 1993, \nat, 364, 507

\bibitem[{{Nozawa} {et~al.}(2010){Nozawa}, {Kozasa}, {Tominaga}, {Maeda},
  {Umeda}, {Nomoto}, \& {Krause}}]{nozawa10}
{Nozawa}, T., {Kozasa}, T., {Tominaga}, N., {et~al.} 2010, \apj, 713, 356

\bibitem[{{Oke} {et~al.}(1995){Oke}, {Cohen}, {Carr}, {Cromer}, {Dingizian},
  {Harris}, {Labrecque}, {Lucinio}, {Schaal}, {Epps}, \& {Miller}}]{Oke_1995}
{Oke}, J.~B., {Cohen}, J.~G., {Carr}, M., {et~al.} 1995, \pasp, 107, 375

\bibitem[{{Perley}(2019)}]{Perley_2019}
{Perley}, D.~A. 2019, \pasp, 131, 084503

\bibitem[{{Perrin} {et~al.}(2014){Perrin}, {Sivaramakrishnan}, {Lajoie},
  {Elliott}, {Pueyo}, {Ravindranath}, \& {Albert}}]{Perrin_2014}
{Perrin}, M.~D., {Sivaramakrishnan}, A., {Lajoie}, C.-P., {et~al.} 2014, in
  Society of Photo-Optical Instrumentation Engineers (SPIE) Conference Series,
  Vol. 9143, Space Telescopes and Instrumentation 2014: Optical, Infrared, and
  Millimeter Wave, ed. J.~{Oschmann}, Jacobus~M., M.~{Clampin}, G.~G. {Fazio},
  \& H.~A. {MacEwen}, 91433X

\bibitem[{Pierel(2024)}]{pierel_2024_12100100}
Pierel, J. 2024, {Space-Phot: Simple Python-Based Photometry for Space
  Telescopes, Zenodo 12100100}

\bibitem[{{Priestley} {et~al.}(2022){Priestley}, {Arias}, {Barlow}, \& {De
  Looze}}]{Priestley_2022}
{Priestley}, F.~D., {Arias}, M., {Barlow}, M.~J., \& {De Looze}, I. 2022,
  \mnras, 509, 3163

\bibitem[{{Priestley} {et~al.}(2019){Priestley}, {Barlow}, \& {De
  Looze}}]{Priestley_2019}
{Priestley}, F.~D., {Barlow}, M.~J., \& {De Looze}, I. 2019, \mnras, 485, 440

\bibitem[{{Ressler} {et~al.}(2015){Ressler}, {Sukhatme}, {Franklin}, {Mahoney},
  {Thelen}, {Bouchet}, {Colbert}, {Cracraft}, {Dicken}, {Gastaud}, {Goodson},
  {Eccleston}, {Moreau}, {Rieke}, \& {Schneider}}]{Ressler_2015}
{Ressler}, M.~E., {Sukhatme}, K.~G., {Franklin}, B.~R., {et~al.} 2015, \pasp,
  127, 675

\bibitem[{Rest {et~al.}(2023)Rest, Pierel, Correnti, Canipe, Hilbert, Engesser,
  Sunnquist, \& Fox}]{Rest_2023}
Rest, A., Pierel, J., Correnti, M., {et~al.} 2023, arminrest/jhat: The JWST HST
  Alignment Tool (JHAT)

\bibitem[{{Richardson} {et~al.}(2006){Richardson}, {Branch}, \&
  {Baron}}]{richardson06}
{Richardson}, D., {Branch}, D., \& {Baron}, E. 2006, \aj, 131, 2233

\bibitem[{{Rieke} \& {Wright}(2022)}]{Rieke_2022}
{Rieke}, G. \& {Wright}, G. 2022, Nature Astronomy, 6, 891

\bibitem[{{Rieke} {et~al.}(2015){Rieke}, {Wright}, {B{\"o}ker}, {Bouwman},
  {Colina}, {Glasse}, {Gordon}, {Greene}, {G{\"u}del}, {Henning}, {Justtanont},
  {Lagage}, {Meixner}, {N{\o}rgaard-Nielsen}, {Ray}, {Ressler}, {van Dishoeck},
  \& {Waelkens}}]{Rieke_2015}
{Rieke}, G.~H., {Wright}, G.~S., {B{\"o}ker}, T., {et~al.} 2015, \pasp, 127,
  584

\bibitem[{{Sarangi}(2022)}]{sarangi22}
{Sarangi}, A. 2022, \aap, 668, A57

\bibitem[{{Seitenzahl} {et~al.}(2014){Seitenzahl}, {Timmes}, \&
  {Magkotsios}}]{Seitenzahl14}
{Seitenzahl}, I.~R., {Timmes}, F.~X., \& {Magkotsios}, G. 2014, \apj, 792, 10

\bibitem[{{Shahbandeh} {et~al.}(2023){Shahbandeh}, {Sarangi}, {Temim},
  {Szalai}, {Fox}, {Tinyanont}, {Dwek}, {Dessart}, {Filippenko}, {Brink},
  {Foley}, {Jencson}, {Pierel}, {Zs{\'\i}ros}, {Rest}, {Zheng}, {Andrews},
  {Clayton}, {De}, {Engesser}, {Gezari}, {Gomez}, {Gonzaga}, {Johansson},
  {Kasliwal}, {Lau}, {De Looze}, {Marston}, {Milisavljevic}, {O'Steen},
  {Siebert}, {Skrutskie}, {Smith}, {Strolger}, {Van Dyk}, {Wang}, {Williams},
  {Williams}, {Xiao}, \& {Yang}}]{Shahbandeh2023}
{Shahbandeh}, M., {Sarangi}, A., {Temim}, T., {et~al.} 2023, \mnras, 523, 6048

\bibitem[{{Sibthorpe} {et~al.}(2010){Sibthorpe}, {Ade}, {Bock}, {Chapin},
  {Devlin}, {Dicker}, {Griffin}, {Gundersen}, {Halpern}, {Hargrave}, {Hughes},
  {Jeong}, {Kaneda}, {Klein}, {Koo}, {Lee}, {Marsden}, {Martin}, {Mauskopf},
  {Moon}, {Netterfield}, {Olmi}, {Pascale}, {Patanchon}, {Rex}, {Roy}, {Scott},
  {Semisch}, {Truch}, {Tucker}, {Tucker}, {Viero}, \& {Wiebe}}]{sibthorpe10}
{Sibthorpe}, B., {Ade}, P.~A.~R., {Bock}, J.~J., {et~al.} 2010, The
  Astrophysical Journal, 719, 1553

\bibitem[{{Smith} {et~al.}(2017){Smith}, {Kilpatrick}, {Mauerhan}, {Andrews},
  {Margutti}, {Fong}, {Graham}, {Zheng}, {Kelly}, {Filippenko}, \&
  {Fox}}]{smith17}
{Smith}, N., {Kilpatrick}, C.~D., {Mauerhan}, J.~C., {et~al.} 2017, \mnras,
  466, 3021

\bibitem[{{Smith} {et~al.}(2011){Smith}, {Li}, {Filippenko}, \&
  {Chornock}}]{Smith_2011}
{Smith}, N., {Li}, W., {Filippenko}, A.~V., \& {Chornock}, R. 2011, \mnras,
  412, 1522

\bibitem[{{Sugerman} {et~al.}(2012){Sugerman}, {Andrews}, {Barlow}, {Clayton},
  {Ercolano}, {Ghavamian}, {Kennicutt}, {Krause}, {Meixner}, \&
  {Otsuka}}]{sugerman12}
{Sugerman}, B. E.~K., {Andrews}, J.~E., {Barlow}, M.~J., {et~al.} 2012, \apj,
  749, 170

\bibitem[{{Sugerman} \& {Crotts}(2002)}]{sugerman02}
{Sugerman}, B. E.~K. \& {Crotts}, A. P.~S. 2002, \apjl, 581, L97

\bibitem[{{Szalai} {et~al.}(2021){Szalai}, {Fox}, {Arendt}, {Dwek}, {Andrews},
  {Clayton}, {Filippenko}, {Johansson}, {Kelly}, {Krafton}, {Marston},
  {Mauerhan}, \& {Van Dyk}}]{szalai21}
{Szalai}, T., {Fox}, O.~D., {Arendt}, R.~G., {et~al.} 2021, \apj, 919, 17

\bibitem[{{Szalai} \& {Vink{\'o}}(2013)}]{szalai13}
{Szalai}, T. \& {Vink{\'o}}, J. 2013, \aap, 549, A79

\bibitem[{{Szalai} {et~al.}(2011){Szalai}, {Vink{\'o}}, {Balog},
  {G{\'a}sp{\'a}r}, {Block}, \& {Kiss}}]{szalai11}
{Szalai}, T., {Vink{\'o}}, J., {Balog}, Z., {et~al.} 2011, \aap, 527, A61

\bibitem[{{Szalai} {et~al.}(2019){Szalai}, {Zs{\'\i}ros}, {Fox}, {Pejcha}, \&
  {M{\"u}ller}}]{szalai19}
{Szalai}, T., {Zs{\'\i}ros}, S., {Fox}, O.~D., {Pejcha}, O., \& {M{\"u}ller},
  T. 2019, \apjs, 241, 38

\bibitem[{{Tanaka} {et~al.}(2012){Tanaka}, {Nozawa}, {Sakon}, {Onaka},
  {Arimatsu}, {Ohsawa}, {Maeda}, {Wada}, {Matsuhara}, \& {Kaneda}}]{tanaka12}
{Tanaka}, M., {Nozawa}, T., {Sakon}, I., {et~al.} 2012, \apj, 749, 173

\bibitem[{{Temim} \& {Dwek}(2013)}]{temim13}
{Temim}, T. \& {Dwek}, E. 2013, ApJ, 774, 8

\bibitem[{{Tinyanont} {et~al.}(2016){Tinyanont}, {Kasliwal}, {Fox}, {Lau},
  {Smith}, {Williams}, {Jencson}, {Perley}, {Dykhoff}, {Gehrz}, {Johansson},
  {Van Dyk}, {Masci}, {Cody}, \& {Prince}}]{tinyanont16}
{Tinyanont}, S., {Kasliwal}, M.~M., {Fox}, O.~D., {et~al.} 2016, \apj, 833, 231

\bibitem[{{Van Dyk}(2013)}]{vandyk13}
{Van Dyk}, S.~D. 2013, \aj, 145, 118

\bibitem[{{Van Dyk} {et~al.}(2002){Van Dyk}, {Garnavich}, {Filippenko},
  {H{\"o}flich}, {Kirshner}, {Kurucz}, \& {Challis}}]{vandyk02}
{Van Dyk}, S.~D., {Garnavich}, P.~M., {Filippenko}, A.~V., {et~al.} 2002,
  \pasp, 114, 1322

\bibitem[{{Vasiliev} \& {Shchekinov}(2024)}]{Vasiliev_2024}
{Vasiliev}, E.~O. \& {Shchekinov}, Y.~A. 2024, \mnras, 527, 8755

\bibitem[{{Weiler} {et~al.}(2007){Weiler}, {Williams}, {Panagia}, {Stockdale},
  {Kelley}, {Sramek}, {Van Dyk}, \& {Marcaide}}]{weiler07}
{Weiler}, K.~W., {Williams}, C.~L., {Panagia}, N., {et~al.} 2007, \apj, 671,
  1959

\bibitem[{{Zs{\'\i}ros} {et~al.}(2022){Zs{\'\i}ros}, {Nagy}, \&
  {Szalai}}]{zsiros22}
{Zs{\'\i}ros}, S., {Nagy}, A.~P., \& {Szalai}, T. 2022, \mnras, 509, 3235

\bibitem[{{Zs{\'\i}ros} {et~al.}(2024){Zs{\'\i}ros}, {Szalai}, {De Looze},
  {Sarangi}, {Shahbandeh}, {Fox}, {Temim}, {Milisavljevic}, {Van Dyk}, {Smith},
  {Filippenko}, {Brink}, {Zheng}, {Dessart}, {Jencson}, {Johansson}, {Pierel},
  {Rest}, {Tinyanont}, {Niculescu-Duvaz}, {Barlow}, {Wesson}, {Andrews},
  {Clayton}, {De}, {Dwek}, {Engesser}, {Foley}, {Gezari}, {Gomez}, {Gonzaga},
  {Kasliwal}, {Lau}, {Marston}, {O'Steen}, {Siebert}, {Skrutskie}, {Strolger},
  {Wang}, {Williams}, {Williams}, \& {Xiao}}]{Zsiros_2024}
{Zs{\'\i}ros}, S., {Szalai}, T., {De Looze}, I., {et~al.} 2024, \mnras, 529,
  155

\bibitem[{{Zubko} {et~al.}(2004){Zubko}, {Dwek}, \& {Arendt}}]{Zubko_2004}
{Zubko}, V., {Dwek}, E., \& {Arendt}, R.~G. 2004, \apjs, 152, 211

\end{thebibliography}

\begin{appendix} 
\section{Modeling the JWST/MIRI SED of SN~1993J assuming large dust grains}\label{sec:appendixA}

\begin{figure}[!h]
\centering
    \includegraphics[width=0.24\textwidth]{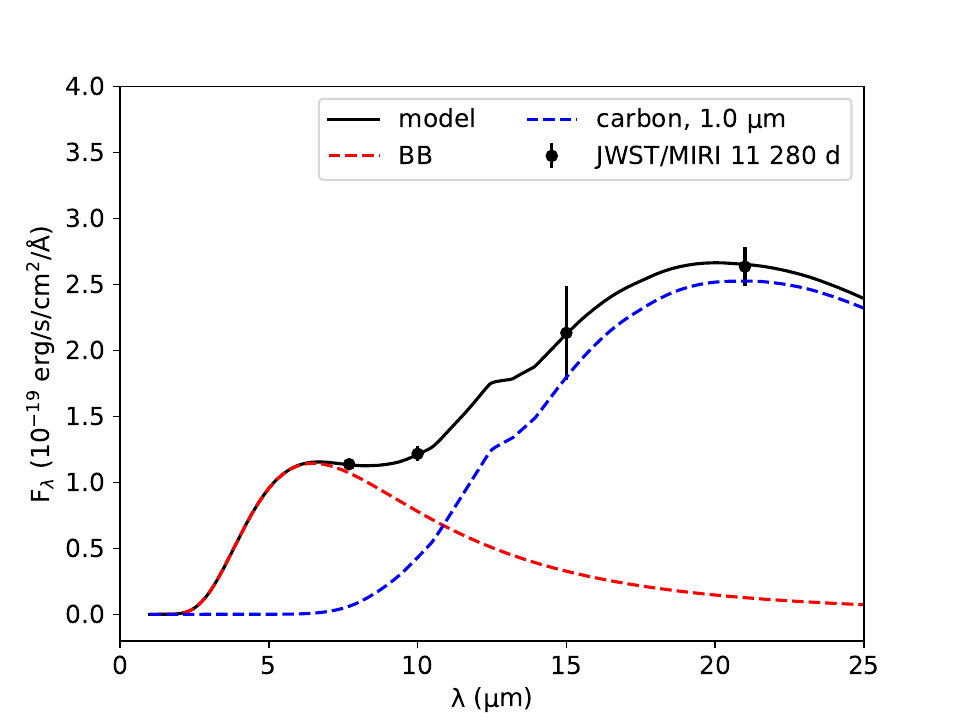}
    \includegraphics[width=0.24\textwidth]{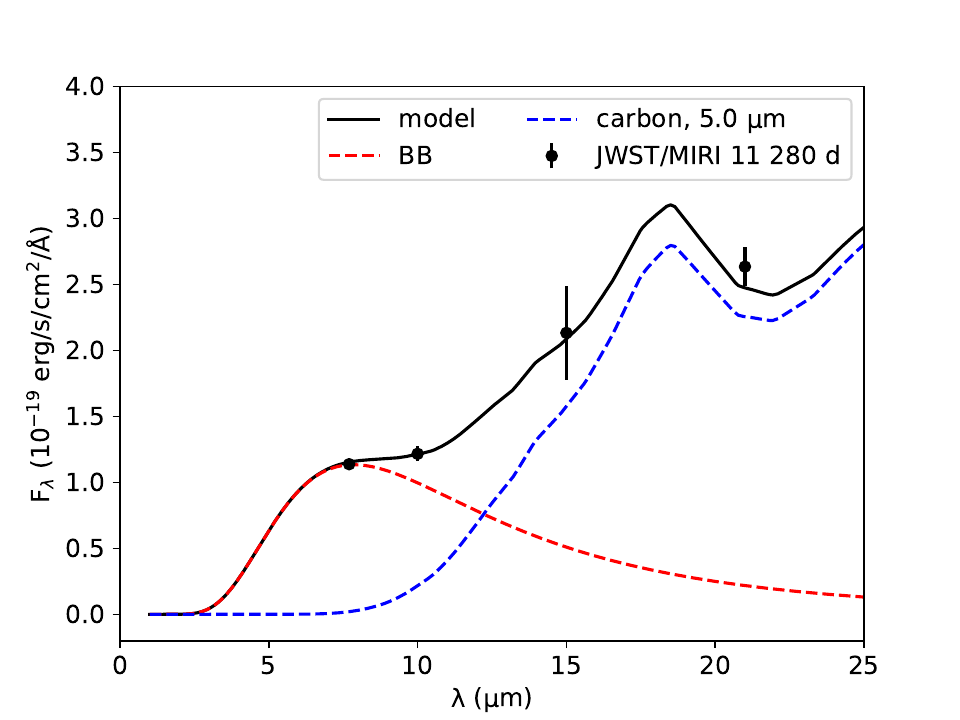}
     \includegraphics[width=0.24\textwidth]{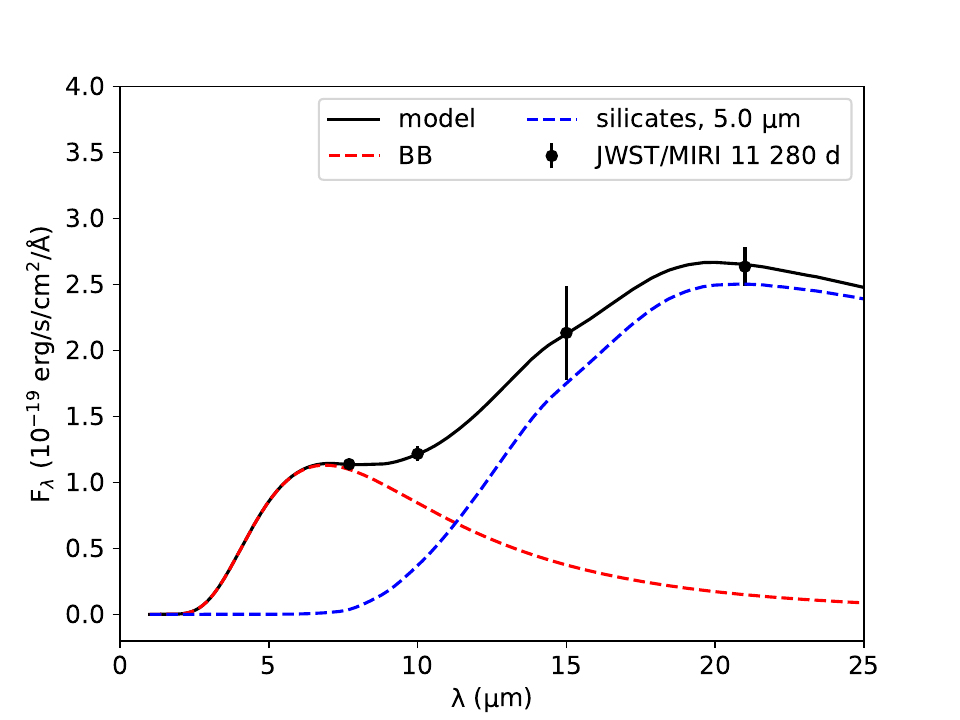}
      \includegraphics[width=0.24\textwidth]{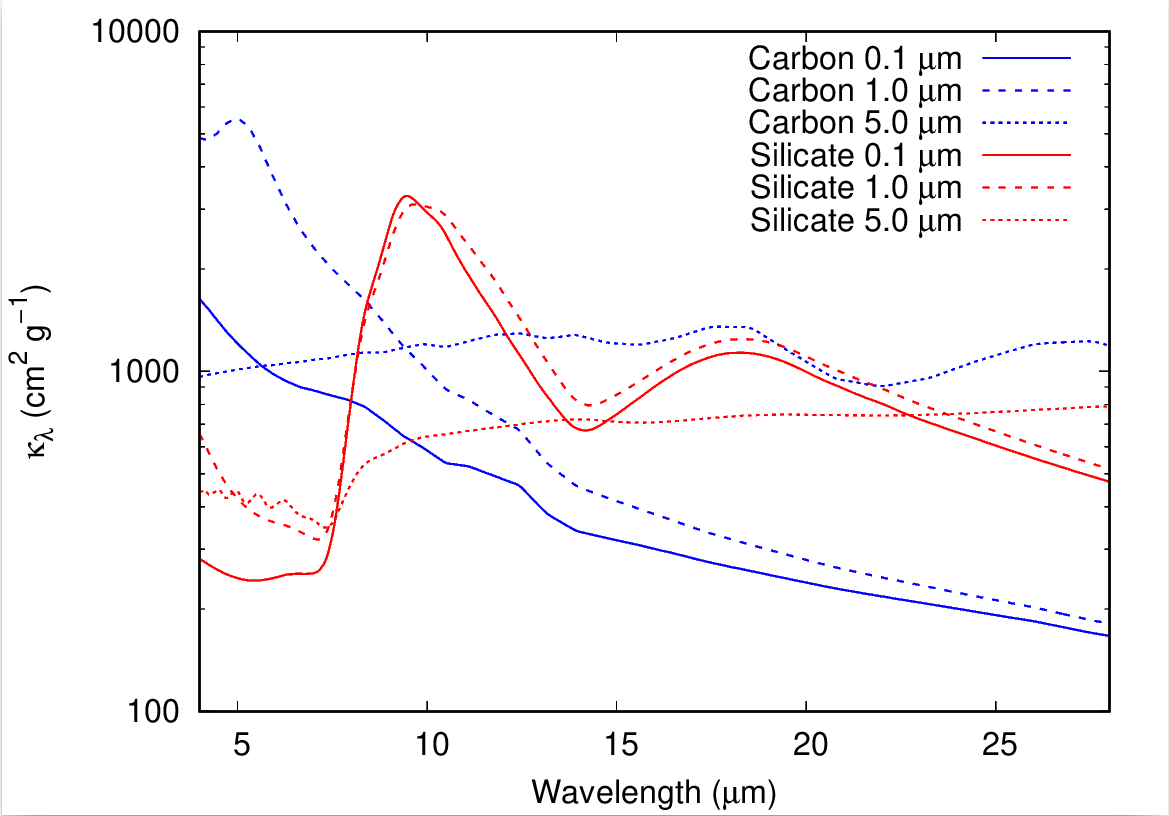}
    \caption{Best-fit two-component models assuming a hot BB and a cold dust component ({\it top left:} amorphous carbon, $a$=1.0 $\mu$m; {\it top right:} amorphous carbon, $a$=5.0 $\mu$m; {\it bottom left:} silicate, $a$=5.0 $\mu$m) and the adopted $\kappa_{\lambda}$ values ({\it bottom right}).}
    \label{fig:SED_large_grains}
\end{figure}

In Fig. \ref{fig:SED_large_grains}, we present the best-fit two-component models assuming a hot BB and a cold dust component consists of large grains ($a$=1.0 and 5.0 $\mu$m amorphous carbon, and $a$=5.0 $\mu$m  silicate dust, respectively), together with the adopted $\kappa_{\lambda}$ curves. We applied this set of fittings only to the average values of the observed JWST/MIRI fluxes.
Models using $a$=1.0 amC dust result in very similar parameters to that of the basic set we used ($a$=0.1 amC dust, see Table \ref{tab:dustpar}), while both $a$=5.0 $\mu$m amC and silicate dust models also result in similar cold dust and hot BB temperatures ($T_\textrm{cold} \sim 110-140$ K, and $T_\textrm{BB,warm} \sim 370-420$ K, respectively), but in smaller cold dust masses ($\sim10^{-3}$ \msolar). 
Nevertheless, as we described above, such large grains are not expected in SN~1993J at this phase and, furthermore, this dataset does not allow us to truly disentangle models with different grain-size distributions.

\section{PSF residuals in JWST/MIRI images}\label{sec:appendixB}

The two figures we present here refer to the potential local background emission we identified at the SN site as residuals after PSF subtraction (see Sec. \ref{sec:anal_echo}). We first examine the quality of the PSF subtraction for multiple isolated stars to ensure the residual emission is not a result of inaccuracies in the PSF models obtained with \texttt{WebbPSF}. 

\begin{figure}
    \includegraphics[width=\columnwidth]{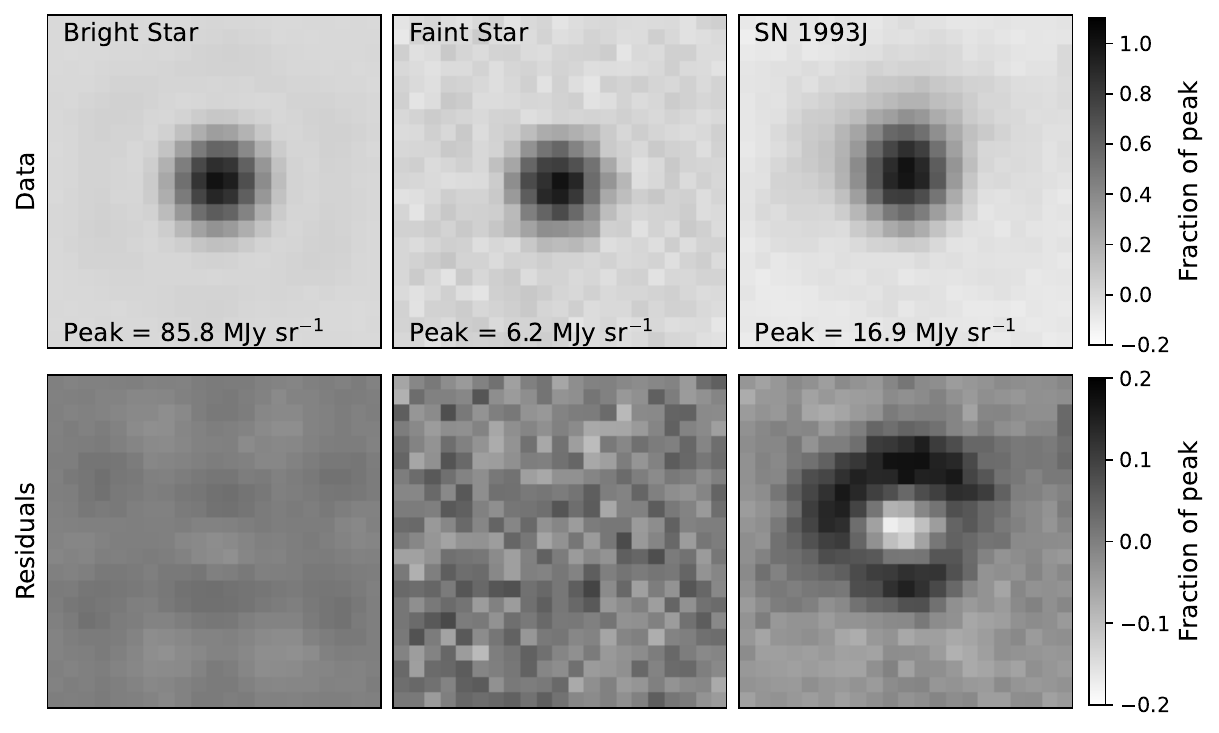}
    \caption{Comparison of subtracting the \texttt{WebbPSF} models from a bright star (left), a faint star (center), and SN~1993J (right) in F1500W images.}
    \label{fig:93J_ring2}
\end{figure}

Figure~\ref{fig:93J_ring2} compares the results of PSF removal from a bright star and a faint star, located at respective positions of ($\alpha$, $\delta$) = ($09^{\rm hr}55^{\rm m}24.86^{\rm s}$, $+69^\circ01'07.2''$) and ($09^{\rm hr}55^{\rm m}23.27^{\rm s}$, $+69^\circ01'01.7''$), along with SN~1993J in the F1500W images. While the bright star shows noticeable PSF residuals, these are $\lesssim 3$\% of the peak of the star and do not show the prominent ``ring'' pattern that is evident in the SN~1993J residuals at up to $\sim 20$\% of the source flux. The faint-star PSF subtracts cleanly, with any PSF residuals below the noise level. This comparison clearly indicates that the residuals around SN~1993J are not due to inaccuracies in the PSF model, but instead arise from real emission in the local SN environment. We note that the ``ring''/doughnut morphology seen here is an artefact of using a large fitting box that results in oversubtraction of the background in the core of the PSF (see Sec.~\ref{sec:anal_echo} for more careful background estimation and PSF fitting).

\begin{figure}
    \includegraphics[width=0.95\columnwidth]{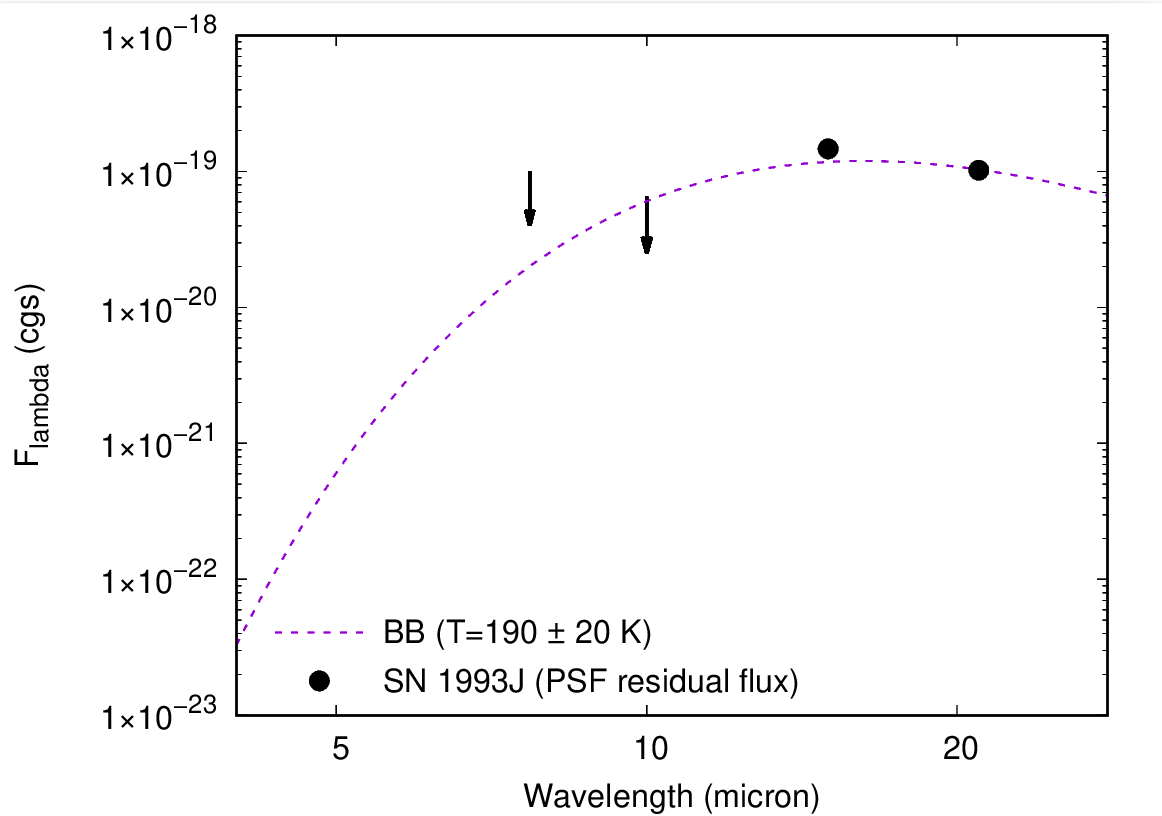}
    \caption{Best-fitting BB curve (dashed purple) to integrated PSF residual fluxes (black circles) and upper limits (black arrows).}
    \label{fig:93J_ring_fit}
\end{figure}

In Fig. \ref{fig:93J_ring_fit}, we show the integrated residual fluxes (F1500W=0.11$\pm$0.01 mJy and F2100W = 0.14$\pm$0.01 mJy, see Sec. \ref{sec:anal_echo}) and upper limits (20~$\mu$Jy for both F770W and F1000W filter images) for the background emission around SN~1993J, together with a fitted BB curve to estimate the temperature of the emitting source. Residual fluxes and upper limits were determined by a simple summation of fluxes within $14 \times 14$ pixel boxes that cover the residual structures of the PSF in the F1500W and F2100W images, but exclude light from nearby point sources. The BB fit suggests a dust temperature of $\sim 190 \pm 20$~K. 

\end{appendix}

\end{document}